\documentclass[twocolumn,prb,amsmath,amssymb,aps,superscriptaddress,floatfix]{revtex4-2}
\usepackage{CJK}
\usepackage{graphicx}
\usepackage{color}
\usepackage{braket}
\usepackage{upgreek}
\usepackage{xr-hyper}
\makeatletter
\newcommand*{\addFileDependency}[1]{
  \typeout{(#1)}
  \@addtofilelist{#1}
  \IfFileExists{#1}{}{\typeout{No file #1.}}
}
\makeatother

\newcommand*{\myexternaldocument}[1]{%
    \externaldocument{#1}%
    \addFileDependency{#1.tex}%
    \addFileDependency{#1.aux}%
}

\myexternaldocument{supp}

\usepackage{dcolumn} 
\usepackage{bm}     
\usepackage{amsfonts}
\usepackage[bookmarks=false,colorlinks,citecolor=blue]{hyperref}
\hypersetup{colorlinks=true,linkcolor=blue,filecolor=blue,citecolor = blue, urlcolor=blue}\usepackage{amsmath}

\usepackage[version=4]{mhchem}
\usepackage{siunitx}

\newcommand{\bfl}{\begin{flushleft}}
\newcommand{\efl}{\end{flushleft}}

\usepackage{multirow}
\usepackage{array}
\newcolumntype{P}[1]{>{\centering\arraybackslash}p{#1}}

\usepackage{subcaption}   
\usepackage{caption}
\begin{document}

\title{Imaging the Superconducting Proximity Effect in S-S'-S Transition Edge Sensors}
\date{\today}

\author{Austin R. Kaczmarek}
\affiliation{Laboratory of Atomic and Solid-State Physics, Cornell University, Ithaca, NY, USA}

\author{Samantha Walker}
\affiliation{Department of Physics, Cornell University, Ithaca, NY, 14853 USA}

\author{Jason Austermann}
\affiliation{National Institute of Standards and Technology, Boulder, CO, 80305 USA}

\author{Douglas Bennett}
\affiliation{National Institute of Standards and Technology, Boulder, CO, 80305 USA}

\author{W. Bertrand Doriese}
\affiliation{National Institute of Standards and Technology, Boulder, CO, 80305 USA}
\author{Shannon M. Duff}
\affiliation{National Institute of Standards and Technology, Boulder, CO, 80305 USA}
\author{Johannes Hubmayr}
\affiliation{National Institute of Standards and Technology, Boulder, CO, 80305 USA}

\author{Kelsey Morgan}
\affiliation{National Institute of Standards and Technology, Boulder, CO, 80305 USA}
\affiliation{Department of Physics, University of Colorado Boulder, Boulder, CO, 80302 USA}

\author{Michael D. Niemack}
\affiliation{Department of Physics, Cornell University, Ithaca, NY, 14853 USA}
\affiliation{Department of Astronomy, Cornell University, Ithaca, NY, 14853}

\author{Dan Schmidt}
\author{Daniel Swetz}
\affiliation{National Institute of Standards and Technology, Boulder, CO, 80305 USA}

\author{Joel Ullom}
\affiliation{National Institute of Standards and Technology, Boulder, CO, 80305 USA}
\affiliation{Department of Physics, University of Colorado Boulder, Boulder, CO, 80302 USA}

\author{Joel Weber}
\affiliation{National Institute of Standards and Technology, Boulder, CO, 80305 USA}
\affiliation{Department of Physics, University of Colorado Boulder, Boulder, CO, 80302 USA}

\author{Katja C. Nowack}
\affiliation{Laboratory of Atomic and Solid-State Physics, Cornell University, Ithaca, NY, USA}
\affiliation{Kavli Institute at Cornell for Nanoscale Science, Ithaca, New York 14853, USA}

\begin{abstract}
Proximity effects at superconducting interfaces, between different superconductors (S-S') or between superconductors and normal metals (S-N), are fundamental to the performance of superconducting electronics, yet only few experiments have directly probed the spatial structure of proximity effects within a device. This is particularly relevant for transition edge sensors (TESs), where the interplay of direct and inverse proximity effects governs detector sensitivity. Here, we use scanning superconducting interference device (SQUID) susceptometry to directly image the local diamagnetic response in functional S-S'-S TES structures. We resolve long range proximity coupling extending over tens of micrometers, revealing that the local transition temperature is dramatically tuned by neighboring regions, being either enhanced by superconducting (S) leads or suppressed by normal metal (N) contacts. Our observations are quantitatively supported by Ginzburg Landau modeling of the device geometry and calculations of the temperature dependent diamagnetism based on self-consistent Usadel equations. By providing spatially resolved measurements of the interplay of proximity effects in TES devices, this work establishes a framework for understanding and controlling superconducting states in heterogeneous superconducting structures.

\end{abstract}

\maketitle

The interface between a superconductor and a normal metal exhibits proximity effects, which modify the electronic properties of both sides \cite{DeGennes1964}. The direct proximity effect induces superconductivity in the metal (N) by the diffusion of Cooper pairs, and the inverse proximity effect simultaneously  weakens the superconducting condensate in the superconductor (S) near the interface due to the normal metal's unpaired electrons. This coupling plays an important role in a broad range of superconducting devices, including Josephson junctions, hybrid quantum circuits, and TESs, and continues to attract interest as a route to engineering new superconducting states by interfacing S with magnetic, topological, or low-dimensional materials \cite{Sharma2022,Liu2024,Blamire2014,Ji2024}.

Proximity effects are also pronounced at the interface of two superconductors, S and S', with different critical temperatures ($T_c > T_c'$). Well above $T_c'$ but below $T_c$, the system behaves like an S-N interface. However, near $T_c'$, the nascent superconductivity in S' can mediate proximity effects over long length scales. The spatial structure of proximity effects in S-S' system has only been studied in a limited number of systems using scanning tunneling microscopy (STM) \cite{Cherkez2014,Kim2012,Wang2022}. These studies probe how the gap evolves across the interface, revealing induced superconductivity that becomes long-ranged near $T_c'$ and  shows a strong dependency on geometry. To our knowledge, no spatially resolved studies currently exist that study the resulting superfluid stiffness or the effect within functional devices. 

Transition edge sensors (TESs) are S-S'-S devices used widely in astronomy for detecting photons spanning millimeter-wave to X-ray wavelengths \cite{Ullom2015,Irwin2005,DeLucia2024}. They exploit the sharp drop in resistance at $T_c'$ to sense minute temperature changes due to the heat from absorbed photons.  Large arrays of TESs have enabled precision measurements of the cosmic microwave background, providing key insights into the universe's structure and evolution \cite{Niemack2010,Filippini2010,Arnold2012,Austermann2012,BICEP22015,Henderson2016,anderson2018,Hui2018,SO2019, Dahal2020,Choi2020}. TESs are typically made from a superconductor with low intrinsic $T_c'=T_{c_0}$ ($\sim \SI{100}{\milli\kelvin}$) combined with a narrow transition width ($\Delta T \sim\ \SI{1}{\milli\kelvin}$ to $\SI{10}{\milli\kelvin}$), and are contacted by leads with a higher transition temperature, $T_{c_L}$, such as Nb ($\sim\SI{9}{\kelvin}$) or Mo ($\sim\SI{1}{\kelvin}$). The TES therefore forms a S-S'-S structure for $T < T_{c_0} < T_{c_L}$ and an S-N-S device when $T_{c_0} < T < T_{c_L}$, where $T$ is the temperature.

Transport studies on various TES designs have shown that device geometry strongly affects the temperature dependent resistance and critical current. These results are often explained by the direct proximity effect from the leads and by the inverse proximity effect from any normal-metal structures incorporated into the device \cite{Sadleir2010,Sadleir2011,Fàbrega20204,de_Wit2020,Nagayoshi2022}. Furthermore, the observation of Fraunhofer-like oscillations in the critical current suggests that some TESs behave as proximitized superconducting weak-links \cite{Sadleir2010,Sadleir2011,Smith2013,Wakeham2018}, reminiscent of Josephson junctions. These findings highlight the significant role of the proximity effect in determining TES behavior, and  suggest opportunities for tunability through careful design. However, exploiting these opportunities requires a detailed understanding of how the geometry changes the local superconducting properties. To date, TESs have primarily been studied using transport measurements which average over the device and obscure non-uniform properties induced by the proximity effect. 

Here, we use scanning superconducting quantum interference device (SQUID) susceptometry to probe the spatial evolution of the superconducting transition in TES devices arising from proximity effects. We measure aluminum-manganese (AlMn) and molybdenum-gold (MoAu) devices across varying geometries that were developed for millimeter-wavelength and X-ray detection applications respectively \cite{Duff2016,Li2016,Weber2020}. Measurements on these functional devices not only provide spatially resolved insights into TES operation for detector engineering but also offer a fundamental advancement by imaging the long-range proximity effect in a way previously unattainable.

\begin{figure*}[htbp]
    \includegraphics[width=1\textwidth]{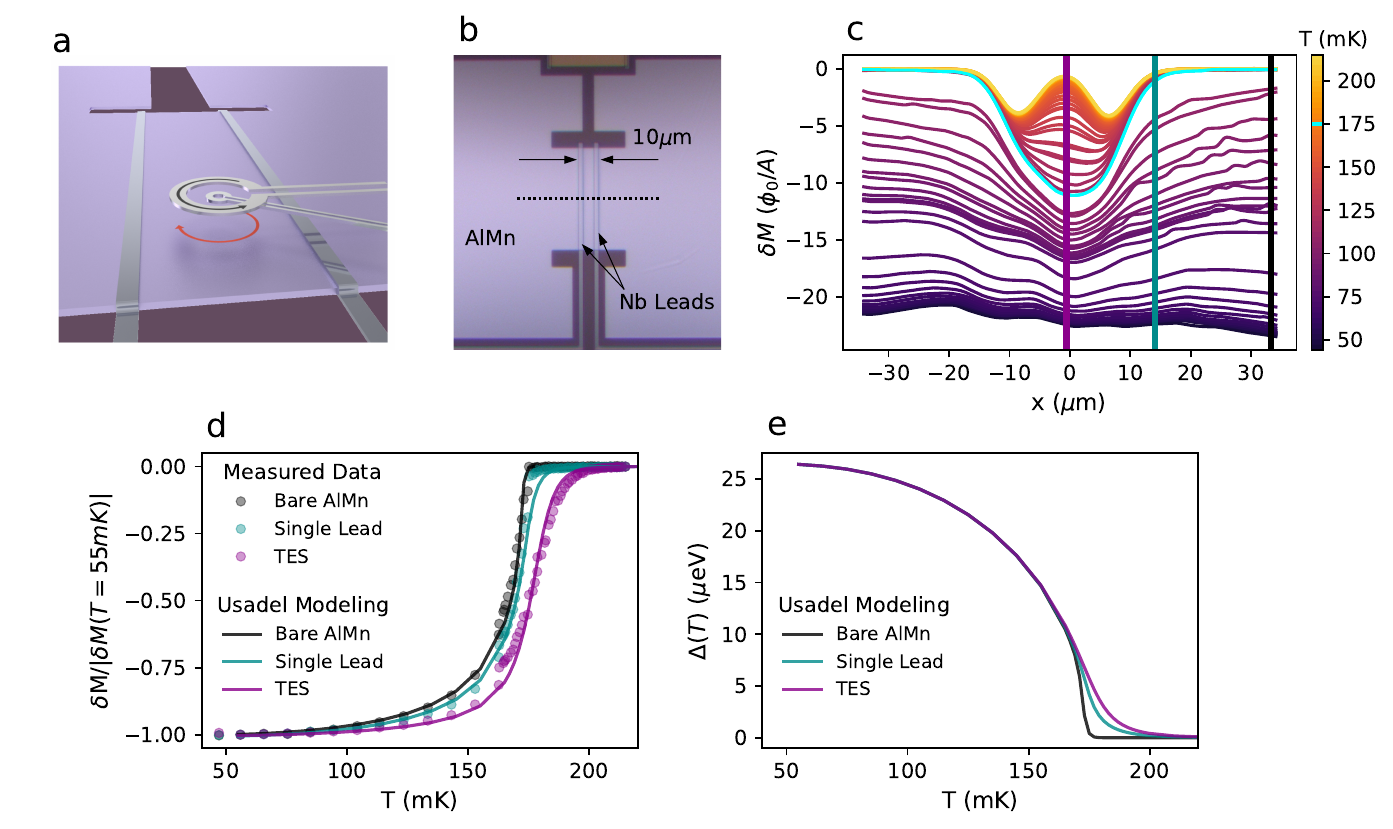}
    \caption{(a) Schematic of the SQUID pickup loop with concentric field coil above the TES. An AC current through the field coil generates a local magnetic field, which is screened by currents in the superconductor. This alters the magnetic flux in the pick-up loop, detected as a change in mutual inductance between the two coils $\delta M$. (b) Optical image of the AlMn TES with Nb leads. AlMn extends to the entire purple colored plane. (c) Horizontal line cuts taken at the center of the device (dashed black line in (b)) of the local DR of the TES as a function of temperature. The aqua colored curve highlights the temperature at which the AlMn between the Nb leads shows an appreciable DR due to the proximity effect, while the bare AlMn far from the leads shows none. (d) DR of TES as a function of temperature for different locations with respect to leads. Points correspond to measured values at fixed positions obtained by taking line cuts through the data in (c) as indicated by the colored vertical lines. Solid lines correspond to a model of the DR obtained from solving a 1D Usadel model of the device. (e) Temperature dependence of superconducting gap obtained from 1D Usadel model of device structure used to calculate the expected DR (solid lines) in (d).}
    \label{AlMn}
\end{figure*}

The scanning SQUID susceptometer consists of two concentric coils: an inner pickup loop connected to the SQUID readout circuit to detect the local magnetic flux, and an outer field coil through which a current is passed to apply a local magnetic field to the sample (see Fig.\ref{AlMn}(a)) \cite{Huber2008}. Far from the sample, current in the field coil couples flux into the pickup loop, giving the bare mutual inductance $M_0$. As the SQUID approaches the sample, currents in superconducting regions at least partially screen the field from the field coil, reducing the mutual inductance $\delta M$. This reduction provides a measure of the local diamagnetic response (DR) and allows us to see superconducting regions. 
The magnitude of the DR is related to the strength and spatial structure of the screening currents and provides a measure of the superconducting penetration depth. All $\delta M$ measurements are performed in zero applied field and bias current.

The first TES device, shown in Fig.\ref{AlMn}(b), consists of a \SI{400}{\nano\meter} thick film of manganese-doped aluminum (AlMn) with Nb leads deposited on top that are \SI{5}{\micro\meter} wide and \SI{200}{\nano\meter} thick. The TES has a length (lead separation) of \SI{10}{\micro\meter}, width of \SI{100}{\micro\meter}, and a normal state resistance of \SI{6.3}{\milli\ohm}. The Nb leads have a nominal transition temperature $T_{c_L}\sim\SI{9}{\kelvin}$, while the bare (non-proximitized) AlMn has a measured $T_{c_0}= \SI{174.5}{\milli\kelvin}$.

Fig.\ref{AlMn}(c) shows the DR ($\delta M$) measured across the center of the TES approximately along the dashed horizontal line shown in Fig.\ref{AlMn}(b). Due to the high aspect ratio of the device, the effect of the edges is negligible near the center of the device, as confirmed by full images of the DR (see Supplemental Section S5). A line cut is therefore sufficient to characterize the device near the center \cite{Walker2025}. For $T_{c_0}\ll T<T_{c_L}$ (yellow curves), we observe only the DR of the Nb leads, which is essentially temperature independent over the range of this measurement, $T\lesssim .02T_{c_L}$. As the temperature decreases, however, the two dips at the lead positions deepen and broaden. This behavior is consistent with proximity-induced superconductivity in the AlMn directly underneath and adjacent to the leads contributing to the DR. At still lower temperatures, the proximitized regions emerging from the two leads extend and merge as is highlighted by the aqua colored curve in Fig.\ref{AlMn}(c), which shows large enhancement of the DR in the center of the TES, while the bare AlMn far from the leads remains normal, demonstrating that the presence of the Nb leads raises the local $T_c$ of the AlMn between them.

To further examine the observed proximity effect, we take slices of the data in Fig.\ref{AlMn}(c) at three fixed positions  indicated by the vertical colored lines: the bare AlMn measured far from the leads (black); at a 5$\mu$m distance from the edge of a single Nb lead outside the active TES area (cyan); and at the center of the TES, 5$\mu$m away from the edges of both Nb leads (magenta). The temperature dependence of the DR $\delta M(T)$ at these locations is shown in Fig.\ref{AlMn}(d). The bare AlMn displays the expected response of a uniform superconductor: a sharp onset of diamagnetism with a finite slope at $T_{c_0}=\SI{174.5}{\milli\kelvin}$, followed by a gradually decreasing slope and saturation upon cooling \cite{Prozorov2006}. 
Near a single lead, the steepest part of the $\delta M$ curve shifts by $\sim\SI{2.5}{\milli\kelvin}$. In addition, the transition is slightly rounded compared to the sharp onset for bare AlMn, shifting the temperature at which $\delta M$ vanishes by approximately \SI{15}{\milli\kelvin}. At the center of the TES, where both leads contribute to the proximity effect in the AlMn, both of these effects are more pronounced: the steepest part of $\delta M(T)$ shifts by $\sim\SI{10}{\milli\kelvin}$, and the additional rounding shifts the onset of diamagnetism by $\sim\SI{35}{\milli\kelvin}$.

A theoretical approach to model the proximity effect in our devices is provided by the Usadel equations, which describe superconductors in the dirty limit where electrons propagate diffusively \cite{Usadel1970}. They have been widely applied to proximitized superconducting systems, capturing the influence of superconductor-metal interfaces \cite{MARTINIS200023,Vodolazov2018,Raj2024}. Unlike Ginzburg-Landau theory, Usadel theory remains quantitatively valid at temperatures far below $T_{c_0}$. Previous work has used the Usadel equations to model the transport characteristics of TESs \cite{Kozorezov2011,MARTINIS200023,Harwin2017}, and found reasonable agreement. However, transport measurements are inherently a global probe and mask the spatial information obtained from solving the equations. Imaging studies using scanning tunneling microscopy (STM) on S-S' and S-S'-S structures have shown that the Usadel equations capture the spatial profile of the proximity effect \cite{Cherkez2014,Kim2012}. In those systems, the effect persists up to a few hundred nm, whereas in the TES devices studied here it extends over much longer length scales, up to tens of \SI{}{\micro\meter}.

We solve the Usadel equations for three 1D geometries to compare with the data in Fig.\ref{AlMn}(d): a bare superconductor, a single S-S' interface, and a S-S'-S structure. This yields the superconducting gap $\Delta(x,T)$ as a function of position and temperature (see Supplemental Section S3). From these solutions, we extract $\Delta(T)$ as shown in Fig.\ref{AlMn}(e) corresponding to the positions marked in Fig.\ref{AlMn}(c,d). Notably, the gap displays rounding near $T_{c_0}$ for the single lead case and even stronger rounding for the case of S' between two leads, reminiscent of the $\delta M(T)$ data shown in Fig.\ref{AlMn}(d). We compare $\Delta(T)$ to $\delta M(T)$ by proceeding in two steps with details provided in   Supplemental Section S3. First, from $\Delta(T)$ we calculate the normalized superfluid density $\rho(T)$ using the measured $T_{c_0}$ of the bare AlMn and assuming a BCS zero temperature gap $\Delta_0 = 1.76k_BT_{c_0}$ \cite{Prozorov2006}. Next, we convert $\rho(T)$ into the expected DR $\delta M(T)$. In this second step, we apply relations derived for a sample with uniform $\rho$ \cite{kirtley2012hsweep}, which means that the model provides only an approximate description of the data. Despite this approximation, the calculated $\delta M(T)$, shown as solid lines in Fig.\ref{AlMn}(d), agrees well with the measured DR in all three cases of a bare superconductor, a S-S' interface, and a S-S'-S device. Different values of the zero temperature penetration depth $\lambda_0$ are required to generate the $\delta M(T)$ model curves for the three cases in Fig.\ref{AlMn}(d), which is reasonable given the spatial averaging of our susceptometry measurement over the small device size (See Supplemental Section S3 for additional discussion).

\begin{figure*}[htbp]
  \centering
    \includegraphics[width=1\textwidth]{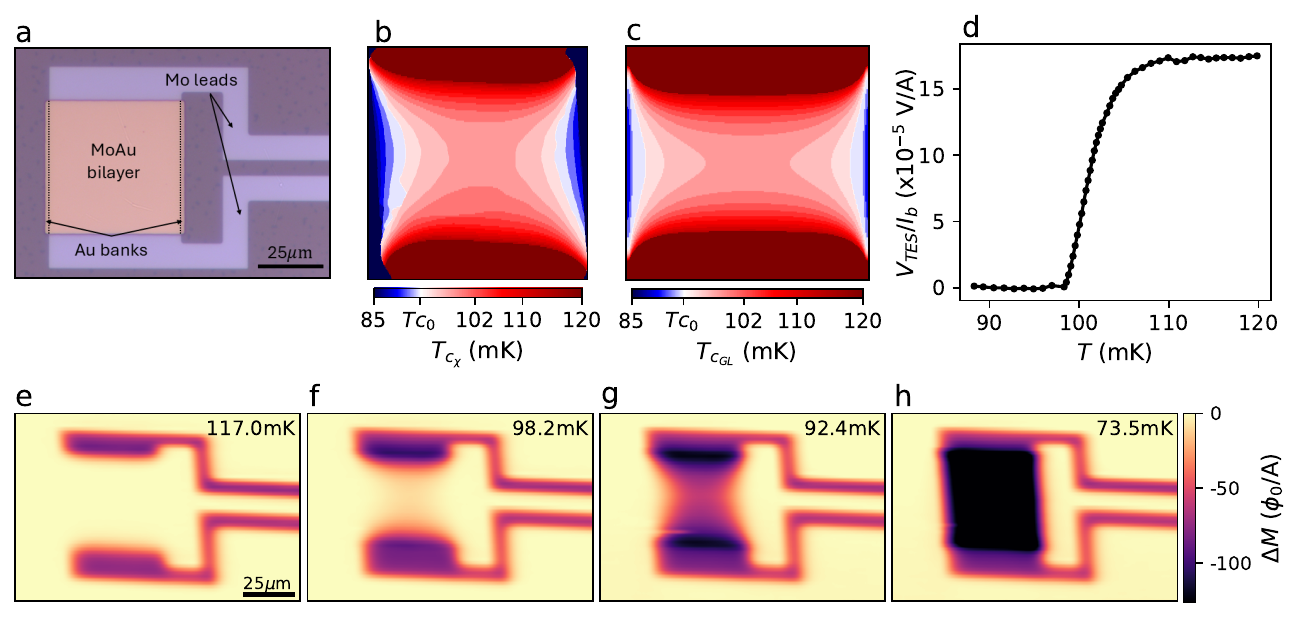}
    \caption{(a) Optical image of the \SI{50}{\micro\meter} TES device structure. Mo leads ($T_{c_L}\sim$\SI{0.9}{\kelvin}) contact the MoAu bilayer with $T_{c_0} =$ \SI{92.8}{\milli\kelvin}. The Au layer overhangs the device forming normal metal banks along the edge of the device. (b) Measured $T_{c_\chi}$ map extracted from a more extensive image series that includes (e-h) by finding the highest temperature at which the local DR surpasses a threshold level above the noise floor. (c) Simulated local $T_{c_{GL}}$ map obtained by solving the GL equations and finding the highest temperature at which the order parameter surpasses a threshold value. Colors of the temperature contours are the same as in (b). (d) Temperature dependent transport of device in a voltage biased circuit with a \SI{180}{\micro\ohm} shunt resistor, plotting ratio of voltage across TES $V_{\mathrm{TES}}$ to total bias current $I_b=$\SI{100}{\micro\ampere}. (e-h) Imaging of the local diamagnetic response of the TES at select temperatures spanning the superconducting transition of the device.     }
    \label{MoAu_50}
\end{figure*}

We next study a TES geometry in which proximity effects produce a spatially varying landscape throughout the full device area. The device structure is shown in Fig.\ref{MoAu_50}(a). A \SI{46}{\nano\meter} Mo bottom layer forms the superconducting leads with $T_{c_L}\sim\SI{0.9}{\kelvin}$, while a \SI{507}{\nano\meter} Au top layer weakens superconductivity in the underlying Mo through the inverse proximity effect. The MoAu bilayer acts as the $`$weak' S' region of the TES \cite{Chen1999,Weber2020}. We determine that $T_{c_0}\sim\SI{92.8}{\milli\kelvin}$ is the intrinsic critical temperature of the bilayer by measuring the temperature dependent DR on a large area on the same chip (see Supplemental Section S6). 
The Au overhangs the Mo edges to prevent superconducting shorts between the leads from non-proximitixed Mo, resulting in normal metal banks in contact with the bilayer. Transport studies of similar TES geometries have explored the influence of the proximity effect from the superconducting leads as well as the inverse proximity effect from the banks \cite{Sadleir2010,Sadleir2011,de_Wit2020,Nagayoshi2022}. We imaged three square TESs of this design with side length \SI{16}{\micro\meter}, \SI{50}{\micro\meter}, and \SI{100}{\micro\meter} fabricated on the same chip. Here we focus on the \SI{50}{\micro\meter} device, which is representative of the others. Images of the other two are provided in Supplemental Section S5.

In Fig.\ref{MoAu_50}(e-h), we show images of the DR of the \SI{50}{\micro\meter} device at several temperatures. At high temperatures $T_{c_0}<<T<T_{c_L}$ (Fig.\ref{MoAu_50}(e)), only the bare Mo leads show a response. As temperature decreases, a DR emerges in the MoAu bilayer near the leads, and gradually spreads inwards, eventually coalescing across the TES in a $`$hourglass' pattern (Fig.\ref{MoAu_50}(f-g)). This spatial distribution reflects the competing effects of the direct proximity from the leads and the inverse proximity from the normal metal Au banks along the TES edges. At much lower temperatures, the entire MoAu bilayer, including the banks, becomes superconducting (Fig.\ref{MoAu_50}(h)) and the devices appears nearly uniform. 

Notably, the DR of the MoAu bilayer at low temperature in Fig.\ref{MoAu_50}(h) is stronger than that of the bare Mo leads, even though both share the same underlying Mo layer. This indicates that the MoAu bilayer does not simply behave as the Mo layer with a reduced $T_c$. Instead, while the Au layer suppresses $T_c$ of Mo through inverse proximity, the Mo layer in-turn proximitizes the Au layer once it becomes superconducting. This contributes additional superfluid in the bilayer, which enhances the DR. This enhancement has been discussed previously in work on superconducting bilayers \cite{Vodolazov2018}.

From an image series of the DR that includes additional temperatures to Fig.\ref{MoAu_50}(e-h) (see Fig.S13 in Supplement), we extract the temperature at which the measured $\delta M$ exceeds a threshold above the noise floor at each position in the device. This defines a spatially varying local transition temperature, $T_{c_\chi}$ shown in Fig.\ref{MoAu_50}(b). 
Compared to the intrinsic $T_{c_0} = 92.8$mK (white) of the bare MoAu bilayer, we find significantly enhanced $T_{c_\chi}$ in the regions near the leads and a suppressed $T_{c_\chi}$ along the edges of the device near the Au banks. This directly reveals the influence of both the direct and inverse proximity effects from the leads and banks. The measured DR, and therefore the extracted $T_{c_\chi}$ map, contains a nontrivial convolution of the local field applied by the field coil, the screening currents induced in the non-uniform sample, and the resulting flux detected by the pickup loop. This  can introduce some distortions compared to the true local $T_{c_\chi}$, for example through detecting faint signals from nearby superconducting regions even though the pickup loop is centered over a non-superconducting region. The primary effect is an effective decrease in the spatial resolution; however, the observed gradual expansion of regions with finite DR extending from the leads directly reflects an underlying growth of the superconducting regions as the device cools.

Next, we compare our imaging to transport measurements of the same device shown in Fig.\ref{MoAu_50}(d). The narrowest hourglass shaped $T_{c_{\chi}}$ contour at the center of the device in Fig.\ref{MoAu_50}(b) corresponds to an image taken at \SI{98.2}{\milli\kelvin}, and marks the highest temperature at which a continuous superconducting path connects the leads. A finite voltage appears at $\sim$ \SI{98.5}{\milli\kelvin}, consistent with the highest temperature identified from the $T_{c_\chi}$ map at which a continuous superconducting path between two leads first develops. The resistive transition is broadened with a clear rounding before reaching a saturated voltage in the normal state. Such rounding has been reported in other TESs and attributed to proximity effects from the leads which gradually reduce the length of the normal region \cite{Sadleir2010}. Our images directly confirm this interpretation. Upon cooling, we observe the induced superconductivity is non-uniform. It emanates from the leads, gradually shrinking the remaining normal region of the device (Fig.\ref{MoAu_50}(f-g)), leading to a corresponding reduction in the resistance. The rounding of the resistive transition becomes more (less) pronounced for smaller (larger) devices as is expected. 

To qualitatively interpret the measured $T_{c_\chi}$ map, we solve the Ginzburg–Landau (GL) equations in the absence of applied fields and currents in 2D for the device geometry \cite{Tinkham,de_Gennes}. We choose the GL framework here because of its lower computational cost compared to the Usadel approach employed above and enables simulations with finer temperature spacing. Although the GL formalism is strictly valid only near $T_{c_0}$, and therefore cannot provide a fully quantitative description as different regions of the device become superconducting at temperatures far from $T_{c_0}$, it remains a useful framework for capturing qualitative features of the onset of superconductivity across the device. We solve the GL equations for the present device geometry at the same temperatures used to image $\delta M$, obtaining the spatial order parameter $\psi(\boldsymbol{x})$ at each temperature. In this, we neglect potential spatial variations across the thickness of the MoAu bilayer. The local transition temperature, $T_{c_{GL}}$, is then defined as the temperature at which $\psi(\boldsymbol{x})$ exceeds a threshold value (see Supplemental Section S4 for details). The resulting $T_{c_{GL}}$ map is shown in Fig.\ref{MoAu_50}(c) and reproduces the key features observed experimentally: enhanced $T_c$ above $T_{c_0}$ near the superconducting leads, suppressed $T_c$ below $T_{c_0}$ near the normal banks, and the characteristic hourglass-shaped contours in the central region of the device.

\begin{figure}[htbp]
  \centering
    \includegraphics[width=.48\textwidth]{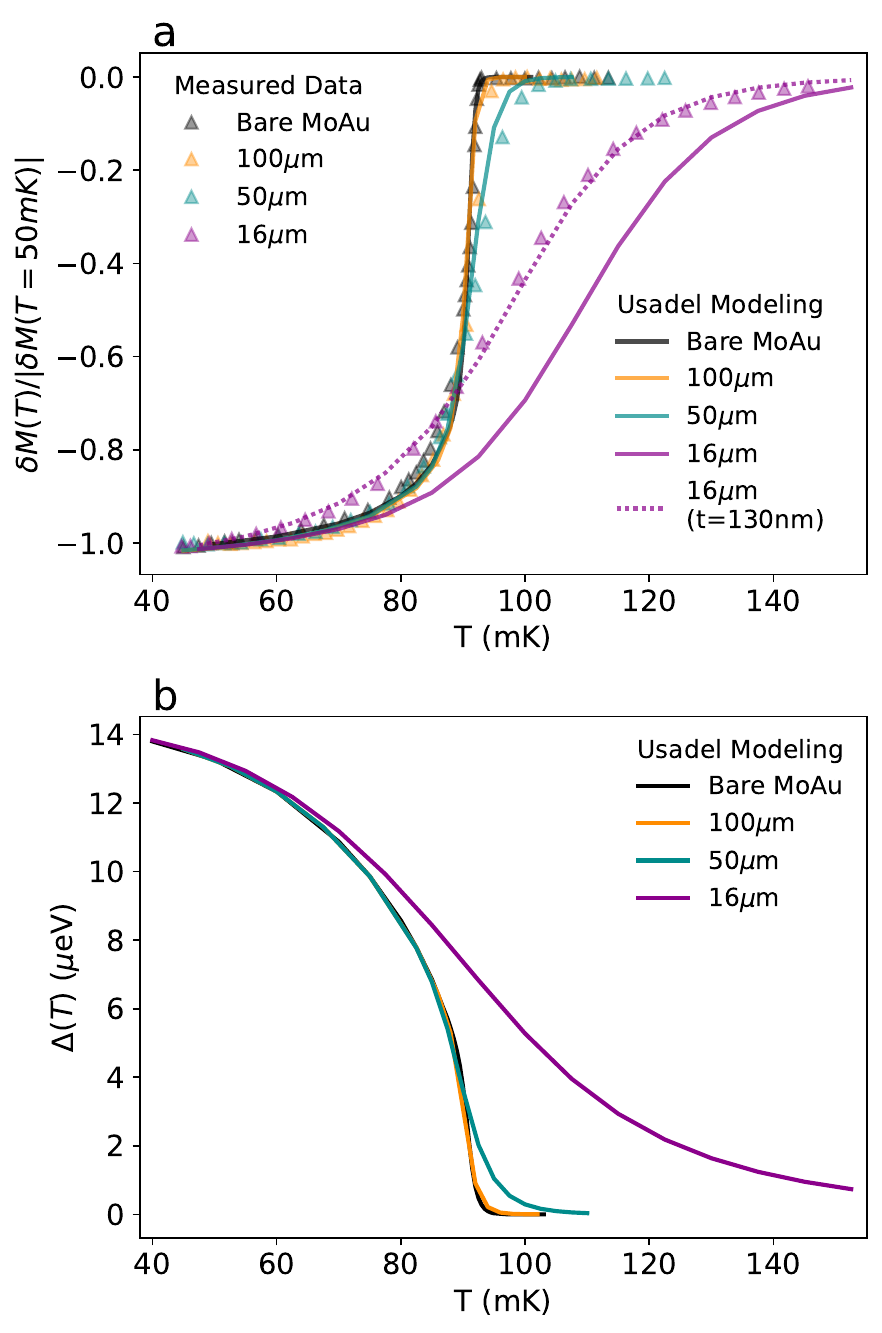}
    \caption{Temperature dependent DR, $\delta M(T)$, for the bare MoAu bilayer and three square TES devices with different dimensions. Data points show the measured $\delta M(T)$ at the center of each TES, rescaled by its value measured at $T=\SI{50}{\milli\kelvin}$. Proximity effects become more pronounced with decreasing device size, as reflected in the shift in $T_{c_{\chi}}$ and the rounding of the transition. Lines in (a) correspond to the DR calculated from a model of the gap at the center of the device as shown in (b) obtained by solving the 2D Usadel equations. Solid lines in (a) assume a common thickness and $\lambda_0$ for all devices, while the dashed line in (a) assumes a smaller effective thickness for the \SI{16}{\micro\meter} device.}
    \label{MoAu_suscep_all}
\end{figure}

\begin{figure*}[htbp]
  \centering
    \includegraphics[width=1\textwidth]{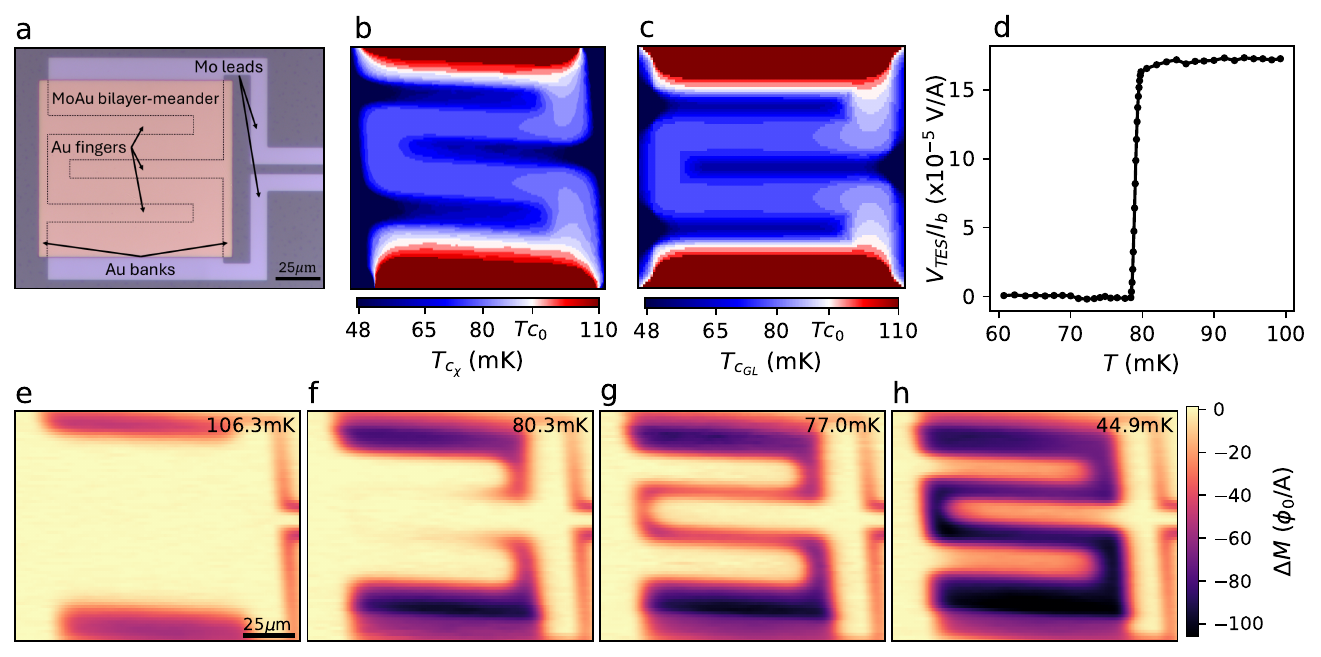}
    \caption{(a) Optical image of the \SI{100}{\micro\meter} meander TES device structure. The Mo layer is patterned into the leads and meander, while the Au layer is deposited on top, forming a MoAu bilayer in the meander region and Au-only banks and fingers. (b) Measured $T_{c_\chi}$ map. (c) Simulated local $T_{c_{GL}}$ map obtained by solving the GL equations. Both maps are extracted using the same procedure as in Fig. \ref{MoAu_50}, and the same GL parameters are used as in Fig.\ref{MoAu_50}(c). Colors of the temperature contours are the same in (b) and (c). (d) Temperature dependent transport of device in a voltage biased circuit with a \SI{180}{\micro\ohm} shunt resistor, plotting ratio of voltage across TES $V_{\mathrm{TES}}$ to total bias current $I_b=$\SI{100}{\micro\ampere}. (e-h) Images of the DR at select temperatures across the superconducting transition.}
    \label{MoAu_meander}
\end{figure*}

Next, we examine the full temperature dependence of the DR, which reflects the temperature dependence of the superfluid stiffness. We compare the response of a bare MoAu bilayer from the same chip with that measured at the centers of three square MoAu TES devices with lateral dimensions \SI{16}{\micro\meter}, \SI{50}{\micro\meter} and \SI{100}{\micro\meter}. Fig.\ref{MoAu_suscep_all}(a) shows the DR $\delta M(T)$ for the bare MoAu bilayer measured in a region without any nearby superconducting or normal metal structures and for the centers of each TES. The bare MoAu bilayer exhibits a sharp onset of the diamagnetism. The center of the \SI{100}{\micro\meter} device behaves very similarly to the bare MoAu bilayer, apart from a small shift in $T_{c_{\chi}}$, indicating only a weak proximity effect at the device center. As the device size decreases, the proximity induced shift in $T_{c_{\chi}}$ becomes more pronounced, the transition broadens over an extended temperature range, and the onset becomes shallow. In the smallest device,  \SI{16}{\micro\meter}, $T_{c_{\chi}}$ (first onset of finite DR) is nearly twice that of $T_{c_0}$.

The systematic change in the onset of the DR with device size is consistent with expectations from the proximity effects, and is captured in the GL calculations discussed above. To evaluate whether the temperature dependence of the DR can be explained by proximity effects, we turn again to the Usadel equations but now in 2D, accounting for the influence of both the superconducting leads and normal metal banks. For each temperature $T$, the model yields a spatially dependent gap $\Delta(\boldsymbol{x})$ for the three geometries. From this, we extract $\Delta(T)$ at the device center, shown in Fig.\ref{MoAu_suscep_all}(b). The $\Delta(T)$ curves from the Usadel calculations reproduce the main qualitative features of the measured $\delta M(T)$, including an increased $T_{c_{\chi}}$ and a more gradual transition extending over a wider temperature range as the device size decreases. We then use $\Delta(T)$ to calculate a model DR of the device following the same procedure and assumptions as for the AlMn TES (see Supplemental Section S3 for details). The solid lines in Fig.\ref{MoAu_suscep_all}(a) show the results of this modeling, with all curves assuming the same sample thickness ($t=$\SI{550}{\nano\meter}) and penetration depth $\lambda_0$. This yields good agreement for all but the \SI{16}{\micro\meter} device. We can, however, obtain good agreement for the \SI{16}{\micro\meter} device by assuming a smaller effective sample thickness ($t=$\SI{130}{\nano\meter}, dashed line). This discrepancy likely reflects the simplified treatment of the MoAu bilayer in the model, which assumes fixed thickness, whereas the effective thickness of the superconducting layer, and thus the number of underlying cooper pairs, likely vary with temperature due to the proximity coupling between the Mo and Au layers. Additional discussion is provided in Supplemental Section S3. 

To further demonstrate and visualize the role of proximity effects, we next study a device geometry that makes these effects even more pronounced. The device structure is shown in Fig\ref{MoAu_meander}(a). This TES has a dimension of \SI{100}{\micro\meter}, and the Mo layer is patterned into a meander within the TES region. After deposition of the Au layer, the device consists of a MoAu bilayer following the meander, Au banks at the device edges, and additional regions where only Au is present (no Mo underneath), referred to as fingers, extending into the device.

In Fig.\ref{MoAu_meander}(e-h), we show images of the DR at several temperatures. Similar to the \SI{50}{\micro\meter} device, the DR initially appears in the leads at high temperature (Fig.\ref{MoAu_meander}(e)). As temperature decreases, diamagnetism first emerges in the MoAu bilayer adjacent to the leads. As temperature is further decreased, superconductivity progressively extends from the leads, but it follows the meander of the MoAu bilayer within the TES (Fig.\ref{MoAu_meander}(f-g)). It continues to spread until the induced superconductivity from the two leads meet in the middle of the vertical segment of the meander at the left of the device (Fig.\ref{MoAu_meander}(g)). This marks the point at which zero resistance would be measured in a transport measurement, and indeed, the temperature corresponding to Fig.\ref{MoAu_meander}(g) coincides closely with the onset of finite voltage in the transport measurement shown in Fig.\ref{MoAu_meander}(d). At even lower temperature, once the entire MoAu bilayer meander is superconducting,  a finite DR also develops within the Au fingers, regions with no Mo layer present. This indicates that at these temperatures, the superconducting MoAu bilayer proximitizes the Au fingers. The DR is suppressed toward the edges of the Au fingers near the Au banks, consistent with an inverse proximity effect from the banks. It is somewhat surprising that the comparatively $`$weak' superconducting bilayer proximitizes the Au fingers which are \SI{10}{\micro\meter} wide. These key features are summarized in the local $T_{c_\chi}$ map shown in Fig.\ref{MoAu_meander}(b) constructed from an extended temperature series of images as those in Fig.\ref{MoAu_meander}(e-h). Notably, $T_{c_{\chi}}$ for much of the meander device is well below $T_{c_0}$ due to the inverse proximity effect from the Au fingers. We have independently confirmed that the full area of the Au fingers becomes superconducting at low temperature by imaging the current distribution in the presence of an applied bias as a function of temperature. These results will be presented in a separate manuscript.

 We model this device geometry using the GL equations to generate a simulated $T_{c_{GL}}$ map (Fig.\ref{MoAu_meander}(c)). Using the same GL parameters as for the \SI{50}{\micro\meter} device (Fig.\ref{MoAu_50}(c)), the model reproduces the key characteristic features of the measured $T_{c_\chi}$ map: the enhancement of $T_{c_{\chi}}$ above $T_{c_0}$ in regions adjacent to the superconducting leads; the $T_{c_{\chi}}$ contours that snake up the MoAu bilayer meander from each lead and eventually merge at the left vertical segment; the finite $T_{c_{\chi}}$ within the Au fingers; and the suppression of $T_{c_{\chi}}$ near the edges of the Au fingers close to the banks. This correspondence illustrates how the device geometry and proximity effects together shape the spatial variation of superconducting properties.

In summary, we have probed the local diamagnetic response of several S-S'-S TES devices and visualized both the direct and inverse proximity effects arising from superconducting and metallic structures. These effects substantially modify the local onset of a diamagnetic response, $T_{c_{\chi}}$, as well as the shape of its temperature dependence. The key features of the measured $T_{c_{\chi}}$ maps are effectively captured by solving the Ginzburg-Landau equations, while the full $\delta M(T)$ behavior, including far below the critical temperature, is well described by self-consistent Usadel equations. Because $T_{c_{\chi}}$ and $\delta M(T)$ influence device characteristics, such as the temperature dependence of the resistance, the critical current, and magnetic field sensitivity, our results show that proximity effects are an important factor in determining TES performance. More broadly, the ability to image and model proximity-induced spatial variations provides a framework for understanding and engineering the superconducting landscape in any device where a superconductor interfaces with another superconductor or a normal metal. Given the ubiquity of proximity and inverse proximity effects in superconducting structures, these insights extend well beyond TESs to a wide range of superconducting sensors, circuits and hybrid quantum devices.

\begin{acknowledgments}
This work was supported by the Air Force Research Laboratory, Project Grant FA9550-21-1-0429. S. Walker acknowledges support by the National Science Foundation Award No. 2503181.
\end{acknowledgments}

\bibliography{bib}

@article{kirtley2012hsweep,
  title={Scanning {SQUID} susceptometry of a paramagnetic superconductor},
  author = {J. R. Kirtley and B. Kalisky and J. A. Bert and C. Bell and M. Kim and Y. Hikita and H. Y. Hwang and J. H. Ngai and Y. Segal and F. J. Walker and C. H. Ahn and K. A. Moler},
  journal={Phys. Rev. B},
  volume={85},
  number={22},
  pages={224518},
  year={2012},
  publisher={APS}
}

@article{Huber2008,
    author = {Huber, Martin E. and Koshnick, Nicholas C. and Bluhm, Hendrik and Archuleta, Leonard J. and Azua, Tommy and Björnsson, Per G. and Gardner, Brian W. and Halloran, Sean T. and Lucero, Erik A. and Moler, Kathryn A.},
    title = "{Gradiometric micro-SQUID susceptometer for scanning measurements of mesoscopic samples}",
    journal = {Rev. Sci.Inst.},
    volume = {79},
    number = {5},
    pages = {053704},
    year = {2008},
    month = {05},
}

@article{Sadleir2010,
  title = {Longitudinal Proximity Effects in Superconducting Transition-Edge Sensors},
  author = {Sadleir, John E. and Smith, Stephen J. and Bandler, Simon R. and Chervenak, James A. and Clem, John R.},
  journal = {Phys. Rev. Lett.},
  volume = {104},
  issue = {4},
  pages = {047003},
  numpages = {4},
  year = {2010},
  month = {Jan},
  publisher = {American Physical Society}}

@article{Sadleir2011,
  title = {Proximity effects and nonequilibrium superconductivity in transition-edge sensors},
  author = {Sadleir, John E. and Smith, Stephen J. and Robinson, Ian K. and Finkbeiner, Fred M. and Chervenak, James A. and Bandler, Simon R. and Eckart, Megan E. and Kilbourne, Caroline A.},
  journal = {Phys. Rev. B},
  volume = {84},
  issue = {18},
  pages = {184502},
  numpages = {9},
  year = {2011},
  month = {Nov},
  publisher = {American Physical Society}}

@article{Wakeham2018,
    title = {Effects of Normal Metal Features on Superconducting Transition-Edge Sensors},
    author = {N. A. Wakeham and J. S. Adams and S. R. Bandler and J. A. Chervenak and
    A. M. Datesman and M. E. Eckart and F. M. Finkbeiner and R. L. Kelley and
    C. A. Kilbourne and A. R. Miniussi and F. S. Porter and J. E. Sadleir and
    K. Sakai and S. J. Smith and E. J. Wassell and W. Yoon},
    journal = {J. Low Temp. Phys.},
    volume = {193},
    pages = {231–240},
    year = {2018}
}

@article{Nagayoshi2022,
    title = {Lateral Inverse Proximity Efect in Ti/Au Transition Edge
Sensors},
    author = {K. Nagayoshi and M. de Wit and E. Taralli and S. Visser, M. L. Ridder and L. Gottardi and H. Akamatsu and D. Vaccaro and M. P. Bruijn and J.‑R. Gao and J. W. A. den Herder},
    journal = {J. Low Temp. Phys.},
    volume = {209},
    pages = {540-547},
    year = {2022}
}

@article{Ullom2015,
year = {2015},
month = {jul},
publisher = {IOP Publishing},
volume = {28},
number = {8},
pages = {084003},
author = {Ullom, Joel N and Bennett, Douglas A},
title = {Review of superconducting transition-edge sensors for x-ray and gamma-ray spectroscopy*},
journal = {Superconductor Science and Technology},
}

@Article{DeLucia2024,
AUTHOR = {De Lucia, Mario and Dal Bo, Paolo and Di Giorgi, Eugenia and Lari, Tommaso and Puglia, Claudio and Paolucci, Federico},
TITLE = {Transition Edge Sensors: Physics and Applications},
JOURNAL = {Instruments},
VOLUME = {8},
YEAR = {2024},
NUMBER = {4},
ARTICLE-NUMBER = {47},
ISSN = {2410-390X},
}

@Inbook{Irwin2005,
author="Irwin, K.D.
and Hilton, G.C.",
editor="Enss, Christian",
title="Transition-Edge Sensors",
bookTitle="Cryogenic Particle Detection",
year="2005",
publisher="Springer Berlin Heidelberg",
address="Berlin, Heidelberg",
pages="63--150",
}

@article{de_Wit2020,
    author = {de Wit, M. and Gottardi, L. and Taralli, E. and Nagayoshi, K. and Ridder, M. L. and Akamatsu, H. and Bruijn, M. P. and D’Andrea, M. and van der Kuur, J. and Ravensberg, K. and Vaccaro, D. and Visser, S. and Gao, J. R. and den Herder, J.-W. A.},
    title = {High aspect ratio transition edge sensors for x-ray spectrometry},
    journal = {Journal of Applied Physics},
    volume = {128},
    number = {22},
    pages = {224501},
    year = {2020},
    month = {12},
    issn = {0021-8979},
}

@article{Fàbrega20204,
    author = {Fàbrega, Lourdes and Camón, Agustín and Pobes, Carlos and Strichovanec, Pavel},
    title = {On the physical origin of the superconducting transition in transition-edge sensors},
    journal = {Journal of Applied Physics},
    volume = {136},
    number = {15},
    pages = {154503},
    year = {2024},
    month = {10},
    issn = {0021-8979},
}

@article{Smith2013,
    author = {Smith, Stephen J. and Adams, Joseph S. and Bailey, Catherine N. and Bandler, Simon R. and Busch, Sarah E. and Chervenak, James A. and Eckart, Megan E. and Finkbeiner, Fred M. and Kilbourne, Caroline A. and Kelley, Richard L. and Lee, Sang-Jun and Porst, Jan-Patrick and Porter, Frederick S. and Sadleir, John E.},
    title = {Implications of weak-link behavior on the performance of Mo/Au bilayer transition-edge sensors},
    journal = {Journal of Applied Physics},
    volume = {114},
    number = {7},
    pages = {074513},
    year = {2013},
    month = {08},
    issn = {0021-8979},
}

@article{Cherkez2014,
  title = {Proximity Effect between Two Superconductors Spatially Resolved by Scanning Tunneling Spectroscopy},
  author = {Cherkez, V. and Cuevas, J. C. and Brun, C. and Cren, T. and M\'enard, G. and Debontridder, F. and Stolyarov, V. S. and Roditchev, D.},
  journal = {Phys. Rev. X},
  volume = {4},
  issue = {1},
  pages = {011033},
  numpages = {13},
  year = {2014},
  month = {Mar},
  publisher = {American Physical Society},
}

@Article{Kim2012,
AUTHOR = {Jungdae Kim and Victor Chua and Gregory A. Fiete and Hyoungdo Nam and Allan H. MacDonald and Chih-Kang Shih},
TITLE = {Visualization of geometric influences on proximity effects in heterogeneous superconductor thin films},
JOURNAL = {Nature Phys.},
VOLUME = {8},
YEAR = {2012},
pages = {464-469},
}

@article{Wang2022,
title = {Anomalous superconducting proximity effect of planar {Pb}–{RhPb$_2$} heterojunctions in the clean limit},
volume = {7},
issn = {2397-4648},
url = {https://www.nature.com/articles/s41535-022-00529-4},
doi = {10.1038/s41535-022-00529-4},
number = {1},
journal = {npj Quantum Materials},
author = {Wang, Rui-Feng and Xiong, Yan-Ling and Zhu, Qun and Ren, Ming-Qiang and Yan, Hang and Song, Can-Li and Ma, Xu-Cun and Xue, Qi-Kun},
year = {2022},
pages = {116},
}

@article{Vodolazov2018,
year = {2018},
volume = {31},
pages = {115004},
author = {D Yu Vodolazov and A Yu Aladyshkin and E E Pestov and S N Vdovichev and S S Ustavshikov and M Yu Levichev and A V Putilov and P A Yunin and A I El'kina and N N Bukharov and A M Klushin},
title = {Peculiar superconducting properties of a thin film superconductor–normal metal bilayer with large ratio of resistivities},
journal = {Superconductor Science and Technology},
}

@article{Harwin2017,
year = {2017},
volume = {30},
pages = {084001},
author = {R C Harwin and D J Goldie and S Withington},
title = {Modelling proximity effects in transition edge sensors to investigate the influence of lateral metal structures},
journal = {Superconductor Science and Technology},
}

@article{MARTINIS200023,
title = {Calculation of {T$_c$} in a normal-superconductor bilayer using the microscopic-based Usadel theory},
journal = {Nuclear Instruments and Methods in Physics Research Section A: Accelerators, Spectrometers, Detectors and Associated Equipment},
volume = {444},
number = {1},
pages = {23-27},
year = {2000},
issn = {0168-9002},
author = {John M Martinis and G.C Hilton and K.D Irwin and D.A Wollman},
}

@ARTICLE{Kozorezov2011,
  author={Kozorezov, A. G. and Golubov, A. A. and Martin, D. D. E. and de Korte, P. A. J. and Lindeman, M. A. and Hijmering, R. A. and Wigmore, J. K.},
  journal={IEEE Transactions on Applied Superconductivity}, 
  title={Microscopic Model of a Transition Edge Sensor as a Weak Link}, 
  year={2011},
  volume={21},
  number={3},
  pages={250-253}
}

@INPROCEEDINGS{Niemack2010,
       author = {{Niemack}, M.~D. and {Ade}, P.~A.~R. and {Aguirre}, J. and {Barrientos}, F. and {Beall}, J.~A. and {Bond}, J.~R. and {Britton}, J. and {Cho}, H.~M. and {Das}, S. and {Devlin}, M.~J. and {Dicker}, S. and {Dunkley}, J. and {D{\"u}nner}, R. and {Fowler}, J.~W. and {Hajian}, A. and {Halpern}, M. and {Hasselfield}, M. and {Hilton}, G.~C. and {Hilton}, M. and {Hubmayr}, J. and {Hughes}, J.~P. and {Infante}, L. and {Irwin}, K.~D. and {Jarosik}, N. and {Klein}, J. and {Kosowsky}, A. and {Marriage}, T.~A. and {McMahon}, J. and {Menanteau}, F. and {Moodley}, K. and {Nibarger}, J.~P. and {Nolta}, M.~R. and {Page}, L.~A. and {Partridge}, B. and {Reese}, E.~D. and {Sievers}, J. and {Spergel}, D.~N. and {Staggs}, S.~T. and {Thornton}, R. and {Tucker}, C. and {Wollack}, E. and {Yoon}, K.~W.},
        title = "{ACTPol: a polarization-sensitive receiver for the Atacama Cosmology Telescope}",
    booktitle = {Millimeter, Submillimeter, and Far-Infrared Detectors and Instrumentation for Astronomy V},
         year = 2010,
       editor = {{Holland}, Wayne S. and {Zmuidzinas}, Jonas},
       series = {Society of Photo-Optical Instrumentation Engineers (SPIE) Conference Series},
       volume = {7741},
        month = jul,
          eid = {77411S},
        pages = {77411S},
          doi = {10.1117/12.857464},
archivePrefix = {arXiv},
       eprint = {1006.5049},
 primaryClass = {astro-ph.IM},
       adsurl = {https://ui.adsabs.harvard.edu/abs/2010SPIE.7741E..1SN}
}

@INPROCEEDINGS{Filippini2010,
       author = {{Filippini}, J.~P. and {Ade}, P.~A.~R. and {Amiri}, M. and {Benton}, S.~J. and {Bihary}, R. and {Bock}, J.~J. and {Bond}, J.~R. and {Bonetti}, J.~A. and {Bryan}, S.~A. and {Burger}, B. and {Chiang}, H.~C. and {Contaldi}, C.~R. and {Crill}, B.~P. and {Dor{\'e}}, O. and {Farhang}, M. and {Fissel}, L.~M. and {Gandilo}, N.~N. and {Golwala}, S.~R. and {Gudmundsson}, J.~E. and {Halpern}, M. and {Hasselfield}, M. and {Hilton}, G. and {Holmes}, W. and {Hristov}, V.~V. and {Irwin}, K.~D. and {Jones}, W.~C. and {Kuo}, C.~L. and {MacTavish}, C.~J. and {Mason}, P.~V. and {Montroy}, T.~E. and {Morford}, T.~A. and {Netterfield}, C.~B. and {O'Dea}, D.~T. and {Rahlin}, A.~S. and {Reintsema}, C.~D. and {Ruhl}, J.~E. and {Runyan}, M.~C. and {Schenker}, M.~A. and {Shariff}, J.~A. and {Soler}, J.~D. and {Trangsrud}, A. and {Tucker}, C. and {Tucker}, R.~S. and {Turner}, A.~D.},
        title = "{SPIDER: a balloon-borne CMB polarimeter for large angular scales}",
    booktitle = {Millimeter, Submillimeter, and Far-Infrared Detectors and Instrumentation for Astronomy V},
         year = 2010,
       editor = {{Holland}, Wayne S. and {Zmuidzinas}, Jonas},
       series = {Society of Photo-Optical Instrumentation Engineers (SPIE) Conference Series},
       volume = {7741},
        month = jul,
          eid = {77411N},
        pages = {77411N},
          doi = {10.1117/12.857720},
        archivePrefix = {arXiv},
       eprint = {1106.2158},
         primaryClass = {astro-ph.CO},
       adsurl = {https://ui.adsabs.harvard.edu/abs/2010SPIE.7741E..1NF}
}

@INPROCEEDINGS{Arnold2012,
       author = {{Arnold}, K. and {Ade}, P.~A.~R. and {Anthony}, A.~E. and {Barron}, D. and {Boettger}, D. and {Borrill}, J. and {Chapman}, S. and {Chinone}, Y. and {Dobbs}, M.~A. and {Errard}, J. and {Fabbian}, G. and {Flanigan}, D. and {Fuller}, G. and {Ghribi}, A. and {Grainger}, W. and {Halverson}, N. and {Hasegawa}, M. and {Hattori}, K. and {Hazumi}, M. and {Holzapfel}, W.~L. and {Howard}, J. and {Hyland}, P. and {Jaffe}, A. and {Keating}, B. and {Kermish}, Z. and {Kisner}, T. and {Le Jeune}, M. and {Lee}, A.~T. and {Linder}, E. and {Lungu}, M. and {Matsuda}, F. and {Matsumura}, T. and {Miller}, N.~J. and {Meng}, X. and {Morii}, H. and {Moyerman}, S. and {Myers}, M.~J. and {Nishino}, H. and {Paar}, H. and {Quealy}, E. and {Reichardt}, C. and {Richards}, P.~L. and {Ross}, C. and {Shimizu}, A. and {Shimmin}, C. and {Shimon}, M. and {Sholl}, M. and {Siritanasak}, P. and {Spieler}, H. and {Stebor}, N. and {Steinbach}, B. and {Stompor}, R. and {Suzuki}, A. and {Tomaru}, T. and {Tucker}, C. and {Zahn}, O.},
        title = "{The bolometric focal plane array of the POLARBEAR CMB experiment}",
    booktitle = {Millimeter, Submillimeter, and Far-Infrared Detectors and Instrumentation for Astronomy VI},
         year = 2012,
       editor = {{Holland}, Wayne S. and {Zmuidzinas}, Jonas},
       series = {Society of Photo-Optical Instrumentation Engineers (SPIE) Conference Series},
       volume = {8452},
        month = sep,
          eid = {84521D},
        pages = {84521D},
          doi = {10.1117/12.927057},
archivePrefix = {arXiv},
       eprint = {1210.7877},
 primaryClass = {astro-ph.IM},
       adsurl = {https://ui.adsabs.harvard.edu/abs/2012SPIE.8452E..1DA}
}

@INPROCEEDINGS{Austermann2012,
       author = {{Austermann}, J.~E. and {Aird}, K.~A. and {Beall}, J.~A. and {Becker}, D. and {Bender}, A. and {Benson}, B.~A. and {Bleem}, L.~E. and {Britton}, J. and {Carlstrom}, J.~E. and {Chang}, C.~L. and {Chiang}, H.~C. and {Cho}, H. -M. and {Crawford}, T.~M. and {Crites}, A.~T. and {Datesman}, A. and {de Haan}, T. and {Dobbs}, M.~A. and {George}, E.~M. and {Halverson}, N.~W. and {Harrington}, N. and {Henning}, J.~W. and {Hilton}, G.~C. and {Holder}, G.~P. and {Holzapfel}, W.~L. and {Hoover}, S. and {Huang}, N. and {Hubmayr}, J. and {Irwin}, K.~D. and {Keisler}, R. and {Kennedy}, J. and {Knox}, L. and {Lee}, A.~T. and {Leitch}, E. and {Li}, D. and {Lueker}, M. and {Marrone}, D.~P. and {McMahon}, J.~J. and {Mehl}, J. and {Meyer}, S.~S. and {Montroy}, T.~E. and {Natoli}, T. and {Nibarger}, J.~P. and {Niemack}, M.~D. and {Novosad}, V. and {Padin}, S. and {Pryke}, C. and {Reichardt}, C.~L. and {Ruhl}, J.~E. and {Saliwanchik}, B.~R. and {Sayre}, J.~T. and {Schaffer}, K.~K. and {Shirokoff}, E. and {Stark}, A.~A. and {Story}, K. and {Vanderlinde}, K. and {Vieira}, J.~D. and {Wang}, G. and {Williamson}, R. and {Yefremenko}, V. and {Yoon}, K.~W. and {Zahn}, O.},
        title = "{SPTpol: an instrument for CMB polarization measurements with the South Pole Telescope}",
    booktitle = {Millimeter, Submillimeter, and Far-Infrared Detectors and Instrumentation for Astronomy VI},
         year = 2012,
       editor = {{Holland}, Wayne S. and {Zmuidzinas}, Jonas},
       series = {Society of Photo-Optical Instrumentation Engineers (SPIE) Conference Series},
       volume = {8452},
        month = sep,
          eid = {84521E},
        pages = {84521E},
          doi = {10.1117/12.927286},
archivePrefix = {arXiv},
       eprint = {1210.4970},
 primaryClass = {astro-ph.IM},
       adsurl = {https://ui.adsabs.harvard.edu/abs/2012SPIE.8452E..1EA}
}

@ARTICLE{BICEP22015,
       author = {{BICEP2 Collaboration} and {Keck Array Collaboration} and {SPIDER Collaboration} and {Ade}, P.~A.~R. and {Aikin}, R.~W. and {Amiri}, M. and {Barkats}, D. and {Benton}, S.~J. and {Bischoff}, C.~A. and {Bock}, J.~J. and {Bonetti}, J.~A. and {Brevik}, J.~A. and {Buder}, I. and {Bullock}, E. and {Chattopadhyay}, G. and {Davis}, G. and {Day}, P.~K. and {Dowell}, C.~D. and {Duband}, L. and {Filippini}, J.~P. and {Fliescher}, S. and {Golwala}, S.~R. and {Halpern}, M. and {Hasselfield}, M. and {Hildebrandt}, S.~R. and {Hilton}, G.~C. and {Hristov}, V. and {Hui}, H. and {Irwin}, K.~D. and {Jones}, W.~C. and {Karkare}, K.~S. and {Kaufman}, J.~P. and {Keating}, B.~G. and {Kefeli}, S. and {Kernasovskiy}, S.~A. and {Kovac}, J.~M. and {Kuo}, C.~L. and {LeDuc}, H.~G. and {Leitch}, E.~M. and {Llombart}, N. and {Lueker}, M. and {Mason}, P. and {Megerian}, K. and {Moncelsi}, L. and {Netterfield}, C.~B. and {Nguyen}, H.~T. and {O'Brient}, R. and {Ogburn}, R.~W., IV and {Orlando}, A. and {Pryke}, C. and {Rahlin}, A.~S. and {Reintsema}, C.~D. and {Richter}, S. and {Runyan}, M.~C. and {Schwarz}, R. and {Sheehy}, C.~D. and {Staniszewski}, Z.~K. and {Sudiwala}, R.~V. and {Teply}, G.~P. and {Tolan}, J.~E. and {Trangsrud}, A. and {Tucker}, R.~S. and {Turner}, A.~D. and {Vieregg}, A.~G. and {Weber}, A. and {Wiebe}, D.~V. and {Wilson}, P. and {Wong}, C.~L. and {Yoon}, K.~W. and {Zmuidzinas}, J.},
        title = {Antenna-coupled TES Bolometers Used in BICEP2, Keck Array, and Spider},
      journal = {The Astrophysical Journal},
         year = 2015,
        month = oct,
       volume = {812},
       number = {2},
          eid = {176},
        pages = {176},
          doi = {10.1088/0004-637X/812/2/176},
archivePrefix = {arXiv},
       eprint = {1502.00619},
 primaryClass = {astro-ph.IM},
       adsurl = {https://ui.adsabs.harvard.edu/abs/2015ApJ...812..176B}
}

@ARTICLE{Henderson2016,
       author = {{Henderson}, S.~W. and {Allison}, R. and {Austermann}, J. and {Baildon}, T. and {Battaglia}, N. and {Beall}, J.~A. and {Becker}, D. and {De Bernardis}, F. and {Bond}, J.~R. and {Calabrese}, E. and {Choi}, S.~K. and {Coughlin}, K.~P. and {Crowley}, K.~T. and {Datta}, R. and {Devlin}, M.~J. and {Duff}, S.~M. and {Dunkley}, J. and {D{\"u}nner}, R. and {van Engelen}, A. and {Gallardo}, P.~A. and {Grace}, E. and {Hasselfield}, M. and {Hills}, F. and {Hilton}, G.~C. and {Hincks}, A.~D. and {Hloẑek}, R. and {Ho}, S.~P. and {Hubmayr}, J. and {Huffenberger}, K. and {Hughes}, J.~P. and {Irwin}, K.~D. and {Koopman}, B.~J. and {Kosowsky}, A.~B. and {Li}, D. and {McMahon}, J. and {Munson}, C. and {Nati}, F. and {Newburgh}, L. and {Niemack}, M.~D. and {Niraula}, P. and {Page}, L.~A. and {Pappas}, C.~G. and {Salatino}, M. and {Schillaci}, A. and {Schmitt}, B.~L. and {Sehgal}, N. and {Sherwin}, B.~D. and {Sievers}, J.~L. and {Simon}, S.~M. and {Spergel}, D.~N. and {Staggs}, S.~T. and {Stevens}, J.~R. and {Thornton}, R. and {Van Lanen}, J. and {Vavagiakis}, E.~M. and {Ward}, J.~T. and {Wollack}, E.~J.},
        title = "{Advanced ACTPol Cryogenic Detector Arrays and Readout}",
      journal = {Journal of Low Temperature Physics},
         year = 2016,
        month = aug,
       volume = {184},
       number = {3-4},
        pages = {772-779},
          doi = {10.1007/s10909-016-1575-z},
archivePrefix = {arXiv},
       eprint = {1510.02809},
 primaryClass = {astro-ph.IM},
       adsurl = {https://ui.adsabs.harvard.edu/abs/2016JLTP..184..772H}
}

@ARTICLE{Anderson2018,
       author = {{Anderson}, A.~J. and {Ade}, P.~A.~R. and {Ahmed}, Z. and
         {Austermann}, J.~E. and {Avva}, J.~S. and {Barry}, P.~S. and
         {Thakur}, R. Basu and {Bender}, A.~N. and {Benson}, B.~A. and
         {Bleem}, L.~E. and {Byrum}, K. and {Carlstrom}, J.~E. and
         {Carter}, F.~W. and {Cecil}, T. and {Chang}, C.~L. and {Cho}, H.~M. and
         {Cliche}, J.~F. and {Crawford}, T.~M. and {Cukierman}, A. and
         {Denison}, E.~V. and {de Haan}, T. and {Ding}, J. and {Dobbs}, M.~A. and
         {Dutcher}, D. and {Everett}, W. and {Foster}, A. and {Gannon}, R.~N. and
         {Gilbert}, A. and {Groh}, J.~C. and {Halverson}, N.~W. and
         {Harke-Hosemann}, A.~H. and {Harrington}, N.~L. and {Henning}, J.~W. and
         {Hilton}, G.~C. and {Holder}, G.~P. and {Holzapfel}, W.~L. and
         {Huang}, N. and {Irwin}, K.~D. and {Jeong}, O.~B. and {Jonas}, M. and
         {Khaire}, T. and {Knox}, L. and {Kofman}, A.~M. and {Korman}, M. and
         {Kubik}, D. and {Kuhlmann}, S. and {Kuklev}, N. and {Kuo}, C.~L. and
         {Lee}, A.~T. and {Leitch}, E.~M. and {Lowitz}, A.~E. and
         {Meyer}, S.~S. and {Michalik}, D. and {Montgomery}, J. and
         {Nadolski}, A. and {Natoli}, T. and {Nguyen}, H. and {Noble}, G.~I. and
         {Novosad}, V. and {Padin}, S. and {Pan}, Z. and {Pearson}, J. and
         {Posada}, C.~M. and {Rahlin}, A. and {Reichardt}, C.~L. and
         {Ruhl}, J.~E. and {Saunders}, L.~J. and {Sayre}, J.~T. and
         {Shirley}, I. and {Shirokoff}, E. and {Smecher}, G. and
         {Sobrin}, J.~A. and {Stark}, A.~A. and {Story}, K.~T. and {Suzuki}, A. and
         {Tang}, Q.~Y. and {Thompson}, K.~L. and {Tucker}, C. and {Vale}, L.~R. and
         {Vanderlinde}, K. and {Vieira}, J.~D. and {Wang}, G. and
         {Whitehorn}, N. and {Yefremenko}, V. and {Yoon}, K.~W. and
         {Young}, M.~R.},
        title = "{SPT-3G: A Multichroic Receiver for the South Pole Telescope}",
      journal = {Journal of Low Temperature Physics},
         year = "2018",
        month = "Dec",
       volume = {193},
       number = {5-6},
        pages = {1057-1065},
          doi = {10.1007/s10909-018-2007-z},
       adsurl = {https://ui.adsabs.harvard.edu/abs/2018JLTP..193.1057A}
}

@INPROCEEDINGS{Hui2018,
       author = {{Hui}, Howard and {Ade}, P.~A.~R. and {Ahmed}, Z. and {Aikin}, R.~W. and {Alexander}, K.~D. and {Barkats}, D. and {Benton}, S.~J. and {Bischoff}, C.~A. and {Bock}, J.~J. and {Bowens-Rubin}, R. and {Brevik}, J.~A. and {Buder}, I. and {Bullock}, E. and {Buza}, V. and {Connors}, J. and {Cornelison}, J. and {Crill}, B.~P. and {Crumrine}, M. and {Dierickx}, M. and {Duband}, L. and {Dvorkin}, C. and {Filippini}, J.~P. and {Fliescher}, S. and {Grayson}, J. and {Hall}, G. and {Halpern}, M. and {Harrison}, S. and {Hildebrandt}, S.~R. and {Hilton}, G.~C. and {Irwin}, K.~D. and {Kang}, J. and {Karkare}, K.~S. and {Karpel}, E. and {Kaufman}, J.~P. and {Keating}, B.~G. and {Kefeli}, S. and {Kernasovskiy}, S.~A. and {Kovac}, J.~M. and {Kuo}, C. -L. and {Lau}, K. and {Larsen}, N.~A. and {Leitch}, E.~M. and {Lueker}, M. and {Megerian}, K.~G. and {Moncelsi}, L. and {Namikawa}, T. and {Netterfield}, C.~B. and {Nguyen}, H.~T. and {O'Brient}, R. and {Ogburn}, R.~W. and {Palladino}, S. and {Pryke}, C. and {Racine}, B. and {Richter}, S. and {Schwarz}, R. and {Schillaci}, A. and {Sheehy}, C.~D. and {Soliman}, A. and {St. Germaine}, T. and {Staniszewski}, Z.~K. and {Steinbach}, B. and {Sudiwala}, R.~V. and {Teply}, G.~P. and {Thompson}, K.~L. and {Tolan}, J.~E. and {Tucker}, C. and {Turner}, A.~D. and {Umilt{\`a}}, C. and {Vieregg}, A.~G. and {Wandui}, A. and {Weber}, A.~C. and {Wiebe}, D.~V. and {Willmert}, J. and {Wong}, C.~L. and {Wu}, W.~L.~K. and {Yang}, E. and {Yoon}, K.~W. and {Zhang}, C.},
        title = "{BICEP Array: a multi-frequency degree-scale CMB polarimeter}",
    booktitle = {Millimeter, Submillimeter, and Far-Infrared Detectors and Instrumentation for Astronomy IX},
         year = 2018,
       editor = {{Zmuidzinas}, Jonas and {Gao}, Jian-Rong},
       series = {Society of Photo-Optical Instrumentation Engineers (SPIE) Conference Series},
       volume = {10708},
        month = jul,
          eid = {1070807},
        pages = {1070807},
          doi = {10.1117/12.2311725},
archivePrefix = {arXiv},
       eprint = {1808.00568},
 primaryClass = {astro-ph.IM},
       adsurl = {https://ui.adsabs.harvard.edu/abs/2018SPIE10708E..07H}
}

@ARTICLE{SO2019,
       author = {{Ade}, P. and {Aguirre}, James and {Ahmed}, Zeeshan and
         {Aiola}, Simone and {Ali}, Aamir and {Alonso}, David and
         {Alvarez}, Marcelo A. and {Arnold}, Kam and {Ashton}, Peter and
         {Austermann}, Jason and {Awan}, Humna and {Baccigalupi}, Carlo and
         {Baildon}, Taylor and {Barron}, Darcy and {Battaglia}, Nick and
         {Battye}, Richard and {Baxter}, Eric and {Bazarko}, Andrew and
         {Beall}, James A. and {Bean}, Rachel and {Beck}, Dominic and
         {Beckman}, Shawn and {Beringue}, Benjamin and {Bianchini}, Federico and
         {Boada}, Steven and {Boettger}, David and {Bond}, J. Richard and
         {Borrill}, Julian and {Brown}, Michael L. and {Bruno}, Sarah Marie and
         {Bryan}, Sean and {Calabrese}, Erminia and {Calafut}, Victoria and
         {Calisse}, Paolo and {Carron}, Julien and {Challinor}, Anthony and
         {Chesmore}, Grace and {Chinone}, Yuji and {Chluba}, Jens and
         {Cho}, Hsiao-Mei Sherry and {Choi}, Steve and {Coppi}, Gabriele and
         {Cothard}, Nicholas F. and {Coughlin}, Kevin and {Crichton}, Devin and
         {Crowley}, Kevin D. and {Crowley}, Kevin T. and {Cukierman}, Ari and
         {D'Ewart}, John M. and {D{\"u}nner}, Rolando and {de Haan}, Tijmen and
         {Devlin}, Mark and {Dicker}, Simon and {Didier}, Joy and {Dobbs}, Matt and
         {Dober}, Bradley and {Duell}, Cody J. and {Duff}, Shannon and
         {Duivenvoorden}, Adri and {Dunkley}, Jo and {Dusatko}, John and
         {Errard}, Josquin and {Fabbian}, Giulio and {Feeney}, Stephen and
         {Ferraro}, Simone and {Flux{\`a}}, Pedro and {Freese}, Katherine and
         {Frisch}, Josef C. and {Frolov}, Andrei and {Fuller}, George and
         {Fuzia}, Brittany and {Galitzki}, Nicholas and {Gallardo}, Patricio A. and
         {Tomas Galvez Ghersi}, Jose and {Gao}, Jiansong and {Gawiser}, Eric and
         {Gerbino}, Martina and {Gluscevic}, Vera and {Goeckner-Wald}, Neil and
         {Golec}, Joseph and {Gordon}, Sam and {Gralla}, Megan and
         {Green}, Daniel and {Grigorian}, Arpi and {Groh}, John and
         {Groppi}, Chris and {Guan}, Yilun and {Gudmundsson}, Jon E. and
         {Han}, Dongwon and {Hargrave}, Peter and {Hasegawa}, Masaya and
         {Hasselfield}, Matthew and {Hattori}, Makoto and {Haynes}, Victor and
         {Hazumi}, Masashi and {He}, Yizhou and {Healy}, Erin and
         {Henderson}, Shawn W. and {Hervias-Caimapo}, Carlos and
         {Hill}, Charles A. and {Hill}, J. Colin and {Hilton}, Gene and
         {Hilton}, Matt and {Hincks}, Adam D. and {Hinshaw}, Gary and
         {Hlo{\v{z}}ek}, Ren{\'e}e and {Ho}, Shirley and {Ho}, Shuay-Pwu Patty and
         {Howe}, Logan and {Huang}, Zhiqi and {Hubmayr}, Johannes and
         {Huffenberger}, Kevin and {Hughes}, John P. and {Ijjas}, Anna and
         {Ikape}, Margaret and {Irwin}, Kent and {Jaffe}, Andrew H. and
         {Jain}, Bhuvnesh and {Jeong}, Oliver and {Kaneko}, Daisuke and
         {Karpel}, Ethan D. and {Katayama}, Nobuhiko and {Keating}, Brian and
         {Kernasovskiy}, Sarah S. and {Keskitalo}, Reijo and {Kisner}, Theodore and
         {Kiuchi}, Kenji and {Klein}, Jeff and {Knowles}, Kenda and
         {Koopman}, Brian and {Kosowsky}, Arthur and
         {Krachmalnicoff}, Nicoletta and {Kuenstner}, Stephen E. and
         {Kuo}, Chao-Lin and {Kusaka}, Akito and {Lashner}, Jacob and
         {Lee}, Adrian and {Lee}, Eunseong and {Leon}, David and
         {Leung}, Jason S. -Y. and {Lewis}, Antony and {Li}, Yaqiong and
         {Li}, Zack and {Limon}, Michele and {Linder}, Eric and
         {Lopez-Caraballo}, Carlos and {Louis}, Thibaut and {Lowry}, Lindsay and
         {Lungu}, Marius and {Madhavacheril}, Mathew and {Mak}, Daisy and
         {Maldonado}, Felipe and {Mani}, Hamdi and {Mates}, Ben and
         {Matsuda}, Frederick and {Maurin}, Lo{\"\i}c and {Mauskopf}, Phil and
         {May}, Andrew and {McCallum}, Nialh and {McKenney}, Chris and
         {McMahon}, Jeff and {Meerburg}, P. Daniel and {Meyers}, Joel and
         {Miller}, Amber and {Mirmelstein}, Mark and {Moodley}, Kavilan and
         {Munchmeyer}, Moritz and {Munson}, Charles and {Naess}, Sigurd and
         {Nati}, Federico and {Navaroli}, Martin and {Newburgh}, Laura and
         {Nguyen}, Ho Nam and {Niemack}, Michael and {Nishino}, Haruki and
         {Orlowski-Scherer}, John and {Page}, Lyman and {Partridge}, Bruce and
         {Peloton}, Julien and {Perrotta}, Francesca and {Piccirillo}, Lucio and
         {Pisano}, Giampaolo and {Poletti}, Davide and {Puddu}, Roberto and
         {Puglisi}, Giuseppe and {Raum}, Chris and {Reichardt}, Christian L. and
         {Remazeilles}, Mathieu and {Rephaeli}, Yoel and {Riechers}, Dominik and
         {Rojas}, Felipe and {Roy}, Anirban and {Sadeh}, Sharon and
         {Sakurai}, Yuki and {Salatino}, Maria and {Sathyanarayana Rao}, Mayuri and
         {Schaan}, Emmanuel and {Schmittfull}, Marcel and {Sehgal}, Neelima and
         {Seibert}, Joseph and {Seljak}, Uros and {Sherwin}, Blake and
         {Shimon}, Meir and {Sierra}, Carlos and {Sievers}, Jonathan and
         {Sikhosana}, Precious and {Silva-Feaver}, Maximiliano and
         {Simon}, Sara M. and {Sinclair}, Adrian and {Siritanasak}, Praween and
         {Smith}, Kendrick and {Smith}, Stephen R. and {Spergel}, David and
         {Staggs}, Suzanne T. and {Stein}, George and {Stevens}, Jason R. and
         {Stompor}, Radek and {Suzuki}, Aritoki and {Tajima}, Osamu and
         {Takakura}, Satoru and {Teply}, Grant and {Thomas}, Daniel B. and
         {Thorne}, Ben and {Thornton}, Robert and {Trac}, Hy and {Tsai}, Calvin and
         {Tucker}, Carole and {Ullom}, Joel and {Vagnozzi}, Sunny and
         {van Engelen}, Alexander and {Van Lanen}, Jeff and
         {Van Winkle}, Daniel D. and {Vavagiakis}, Eve M. and
         {Verg{\`e}s}, Clara and {Vissers}, Michael and {Wagoner}, Kasey and
         {Walker}, Samantha and {Ward}, Jon and {Westbrook}, Ben and
         {Whitehorn}, Nathan and {Williams}, Jason and {Williams}, Joel and
         {Wollack}, Edward J. and {Xu}, Zhilei and {Yu}, Byeonghee and
         {Yu}, Cyndia and {Zago}, Fernando and {Zhang}, Hezi and
         {Zhu}, Ningfeng and {The Simons Observatory collaboration}},
        title = "{The Simons Observatory: science goals and forecasts}",
      journal = {Journal of Cosmology and Astroparticle Physics},
         year = "2019",
        month = "Feb",
       volume = {2019},
       number = {2},
          eid = {056},
        pages = {056}
}

@article{Dahal2020,
	year = 2020,
	month = {jan},
  
	publisher = {Springer Science and Business Media {LLC}
},
  
	volume = {199},
  
	number = {1-2},
  
	pages = {289--297},
  
	author = {S. Dahal and M. Amiri and J. W. Appel and C. L. Bennett and L. Corbett and R. Datta and K. Denis and T. Essinger-Hileman and M. Halpern and K. Helson and G. Hilton and J. Hubmayr and B. Keller and T. Marriage and C. Nunez and M. Petroff and C. Reintsema and K. Rostem and K. U-Yen and E. Wollack},
  
	title = {The {CLASS} 150/220~{GHz} Polarimeter Array: Design, {Assembly}, and {Characterization}},
  
	journal = {Journal of Low Temperature Physics}
}

@Article{Choi2020,
  author        = {{Choi}, Steve K. and {Hasselfield}, Matthew and {Ho}, Shuay-Pwu Patty and {Koopman}, Brian and {Lungu}, Marius and {Abitbol}, Maximilian H. and {Addison}, Graeme E. and {Ade}, Peter A.~R. and {Aiola}, Simone and {Alonso}, David and {Amiri}, Mandana and {Amodeo}, Stefania and {Angile}, Elio and {Austermann}, Jason E. and {Baildon}, Taylor and {Battaglia}, Nick and {Beall}, James A. and {Bean}, Rachel and {Becker}, Daniel T. and {Bond}, J Richard and {Bruno}, Sarah Marie and {Calabrese}, Erminia and {Calafut}, Victoria and {Campusano}, Luis E. and {Carrero}, Felipe and {Chesmore}, Grace E. and {Cho}, Hsiao-mei and {Clark}, Susan E. and {Cothard}, Nicholas F. and {Crichton}, Devin and {Crowley}, Kevin T. and {Darwish}, Omar and {Datta}, Rahul and {Denison}, Edward V. and {Devlin}, Mark J. and {Duell}, Cody J. and {Duff}, Shannon M. and {Duivenvoorden}, Adriaan J. and {Dunkley}, Jo and {D{\"u}nner}, Rolando and {Essinger-Hileman}, Thomas and {Fankhanel}, Max and {Ferraro}, Simone and {Fox}, Anna E. and {Fuzia}, Brittany and {Gallardo}, Patricio A. and {Gluscevic}, Vera and {Golec}, Joseph E. and {Grace}, Emily and {Gralla}, Megan and {Guan}, Yilun and {Hall}, Kirsten and {Halpern}, Mark and {Han}, Dongwon and {Hargrave}, Peter and {Henderson}, Shawn and {Hensley}, Brandon and {Hill}, J. Colin and {Hilton}, Gene C. and {Hilton}, Matt and {Hincks}, Adam D. and {Hlo{\v{z}}ek}, Ren{\'e}e and {Hubmayr}, Johannes and {Huffenberger}, Kevin M. and {Hughes}, John P. and {Infante}, Leopoldo and {Irwin}, Kent and {Jackson}, Rebecca and {Klein}, Jeff and {Knowles}, Kenda and {Kosowsky}, Arthur and {Lakey}, Vincent and {Li}, Dale and {Li}, Yaqiong and {Li}, Zack and {Lokken}, Martine and {Louis}, Thibaut and {MacInnis}, Amanda and {Madhavacheril}, Mathew and {Maldonado}, Felipe and {Mallaby-Kay}, Maya and {Marsden}, Danica and {Maurin}, Lo{\"\i}c and {McMahon}, Jeff and {Menanteau}, Felipe and {Moodley}, Kavilan and {Morton}, Tim and {Naess}, Sigurd and {Namikawa}, Toshiya and {Nati}, Federico and {Newburgh}, Laura and {Nibarger}, John P. and {Nicola}, Andrina and {Niemack}, Michael D. and {Nolta}, Michael R. and {Orlowski-Sherer}, John and {Page}, Lyman A. and {Pappas}, Christine G. and {Partridge}, Bruce and {Phakathi}, Phumlani and {Prince}, Heather and {Puddu}, Roberto and {Qu}, Frank J. and {Rivera}, Jesus and {Robertson}, Naomi and {Rojas}, Felipe and {Salatino}, Maria and {Schaan}, Emmanuel and {Schillaci}, Alessandro and {Schmitt}, Benjamin L. and {Sehgal}, Neelima and {Sherwin}, Blake D. and {Sierra}, Carlos and {Sievers}, Jon and {Sifon}, Cristobal and {Sikhosana}, Precious and {Simon}, Sara and {Spergel}, David N. and {Staggs}, Suzanne T. and {Stevens}, Jason and {Storer}, Emilie and {Sunder}, Dhaneshwar D. and {Switzer}, Eric R. and {Thorne}, Ben and {Thornton}, Robert and {Trac}, Hy and {Treu}, Jesse and {Tucker}, Carole and {Vale}, Leila R. and {Van Engelen}, Alexander and {Van Lanen}, Jeff and {Vavagiakis}, Eve M. and {Wagoner}, Kasey and {Wang}, Yuhan and {Ward}, Jonathan T. and {Wollack}, Edward J. and {Xu}, Zhilei and {Zago}, Fernando and {Zhu}, Ningfeng},
  title         = {{The Atacama Cosmology Telescope: A Measurement of the Cosmic Microwave Background Power Spectra at 98 and 150 GHz}},
  journal       = {arXiv e-prints},
  year          = {2020},
  pages         = {arXiv:2007.07289},
  month         = jul,
  adsurl        = {https://ui.adsabs.harvard.edu/abs/2020arXiv200707289C},
  archiveprefix = {arXiv},
  eid           = {arXiv:2007.07289},
  eprint        = {2007.07289},
  primaryclass  = {astro-ph.CO},
}

@article{Usadel1970,
  title = {Generalized Diffusion Equation for Superconducting Alloys},
  author = {Usadel, Klaus D.},
  journal = {Phys. Rev. Lett.},
  volume = {25},
  issue = {8},
  pages = {507--509},
  numpages = {0},
  year = {1970},
  month = {Aug},
  publisher = {American Physical Society},
}

@article{Raj2024,
  title = {Self-consistent evaluation of proximity and inverse proximity effects with pair-breaking in diffusive superconducting--normal metal junctions},
  author = {Raj, Arpit and Lee, Patrick A. and Fiete, Gregory A.},
  journal = {Phys. Rev. B},
  volume = {110},
  issue = {18},
  pages = {184504},
  numpages = {11},
  year = {2024},
  month = {Nov},
  publisher = {American Physical Society},
 }

@article{Prozorov2006,
  title={Magnetic penetration depth in unconventional
superconductors},
  author={Prozorov, R. and Giannetta, R. W.},
  journal={Supercond. Sci. Technol},
  volume={19},
  pages={R41-R67},
  year={2006},
}

@ARTICLE{Walker2025,
  author={Walker, Samantha and Kaczmarek, Austin and Austermann, Jason and Bennett, Douglas and Duff, Shannon M. and Hubmayr, Johannes and Keller, Ben and Morgan, Kelsey and Murphy, Colin C. and Swetz, Daniel and Ullom, Joel and Niemack, Michael D. and Nowack, Katja C.},
  journal={IEEE Transactions on Applied Superconductivity}, 
  title={Direct Imaging of Transition-Edge Sensors With Scanning SQUID Microscopy}, 
  year={2025},
  volume={35},
  number={5},
  pages={1-6},
}

@ARTICLE{Weber2020,
  author={Joel C Weber and Kelsey M Morgan and Daikang Yan and Christine G Pappas and Abigail L Wessels and Galen C O’Neil and Doug A Bennett and Gene C Hilton and Daniel S Swetz and Joel N Ullom and Daniel R Schmidt},
  journal={Superconductor Science and Technology}, 
  title={Development of a transition-edge sensor bilayer process providing new modalities for critical temperature control}, 
  year={2020},
  volume={33},
  pages={115002},
}

@ARTICLE{Chen1999,
  author={T C Chen and F M Finkbeiner and A Bier and B DiCamillo},
  journal={Superconductor Science and Technology}, 
  title={Molybdenum-gold proximity bilayers as transition edge sensors for microcalorimeters and bolometers}, 
  year={1999},
  volume={12},
  pages={840},
}

@BOOK{de_Gennes,
author = {De Gennes, P. G.},
title = {Superconductivity Of Metals And Alloys},
publisher = {W. A. Benjamin, Inc.},
year = {1966}
}

@BOOK{Tinkham,
author = {M. Tinkham},
title = {Introduction to Superconductivity},
year = {1999},
publisher={Dover Publications Inc.},
edition = {2}
}

@Article{Duff2016,
author={Duff, S. M.
and Austermann, J.
and Beall, J. A.
and Becker, D.
and Datta, R.
and Gallardo, P. A.
and Henderson, S. W.
and Hilton, G. C.
and Ho, S. P.
and Hubmayr, J.
and Koopman, B. J.
and Li, D.
and McMahon, J.
and Nati, F.
and Niemack, M. D.
and Pappas, C. G.
and Salatino, M.
and Schmitt, B. L.
and Simon, S. M.
and Staggs, S. T.
and Stevens, J. R.
and Van Lanen, J.
and Vavagiakis, E. M.
and Ward, J. T.
and Wollack, E. J.},
title={Advanced ACTPol Multichroic Polarimeter Array Fabrication Process for 150 mm Wafers},
journal={Journal of Low Temperature Physics},
year={2016},
month={Aug},
day={01},
volume={184},
number={3},
pages={634-641},
issn={1573-7357},
}

@Article{Li2016,
author={Li, Dale
and Austermann, Jason E.
and Beall, James A.
and Becker, Daniel T.
and Duff, Shannon M.
and Gallardo, Patricio A.
and Henderson, Shawn W.
and Hilton, Gene C.
and Ho, Shuay-Pwu
and Hubmayr, Johannes
and Koopman, Brian J.
and McMahon, Jeffrey J.
and Nati, Federico
and Niemack, Michael D.
and Pappas, Christine G.
and Salatino, Maria
and Schmitt, Benjamin L.
and Simon, Sara M.
and Staggs, Suzanne T.
and Van Lanen, Jeff
and Ward, Jonathan T.
and Wollack, Edward J.},
title={AlMn Transition Edge Sensors for Advanced ACTPol},
journal={Journal of Low Temperature Physics},
year={2016},
month={Jul},
day={01},
volume={184},
number={1},
pages={66-73},
issn={1573-7357},
}

@article{DeGennes1964,
	title = {Boundary Effects in Superconductors},
	volume = {36},
	issn = {0034-6861},
	pages = {225--237},
	number = {1},
	journal = {Rev. Mod. Phys.},
	author = {De Gennes, P. G.},
	year = {1964},
}

@ARTICLE{Sharma2022,
  author={M M Sharma and Prince Sharma and N K Karn and V P S Awana},
  journal={Superconductor Science and Technology}, 
  title={Comprehensive review on topological superconducting materials and interfaces}, 
  year={2022},
  volume={35},
  pages={083003},
}

@article{Liu2024,
author = {Liu, Yichen and Meng, Qingxiao and Mahmoudi, Pezhman and Wang, Ziyi and Zhang, Ji and Yang, Jack and Li, Wenxian and Wang, Danyang and Li, Zhi and Sorrell, Charles C. and Li, Sean},
title = {Advancing Superconductivity with Interface Engineering},
journal = {Advanced Materials},
volume = {36},
number = {42},
pages = {2405009},
year = {2024}
}

@ARTICLE{Blamire2014,
  author={M G Blamire and J W A Robinson},
  journal={J. Phys.: Condens. Matter}, 
  title={The interface between superconductivity and magnetism: understanding and device prospects}, 
  year={2014},
  volume={26},
  pages={453201},
}

@Article{Ji2024,
author={Ji, Haoran
and Liu, Yi
and Ji, Chengcheng
and Wang, Jian},
title={Two-Dimensional and Interface Superconductivity in Crystalline Systems},
journal={Accounts of Materials Research},
year={2024},
month={Oct},
day={25},
volume={5},
number={10},
pages={1146-1157},
}

@article{BISHOPVANHORN2022,
title = {SuperScreen: An open-source package for simulating the magnetic response of two-dimensional superconducting devices},
journal = {Computer Physics Communications},
volume = {280},
pages = {108464},
year = {2022},
issn = {0010-4655},
author = {Logan {Bishop-Van Horn} and Kathryn A. Moler},
}

\end{document}


\title{Supplementary Information for: Imaging the Superconducting Proximity Effect in S-S'-S Transition Edge Sensor}
\maketitle
\tableofcontents
\newcounter{somecounter}

\newpage
\stepcounter{somecounter}
\section{Supplementary Note \thesomecounter: Device details}

\begin{figure}[htbp]
    \includegraphics[width=1\linewidth]{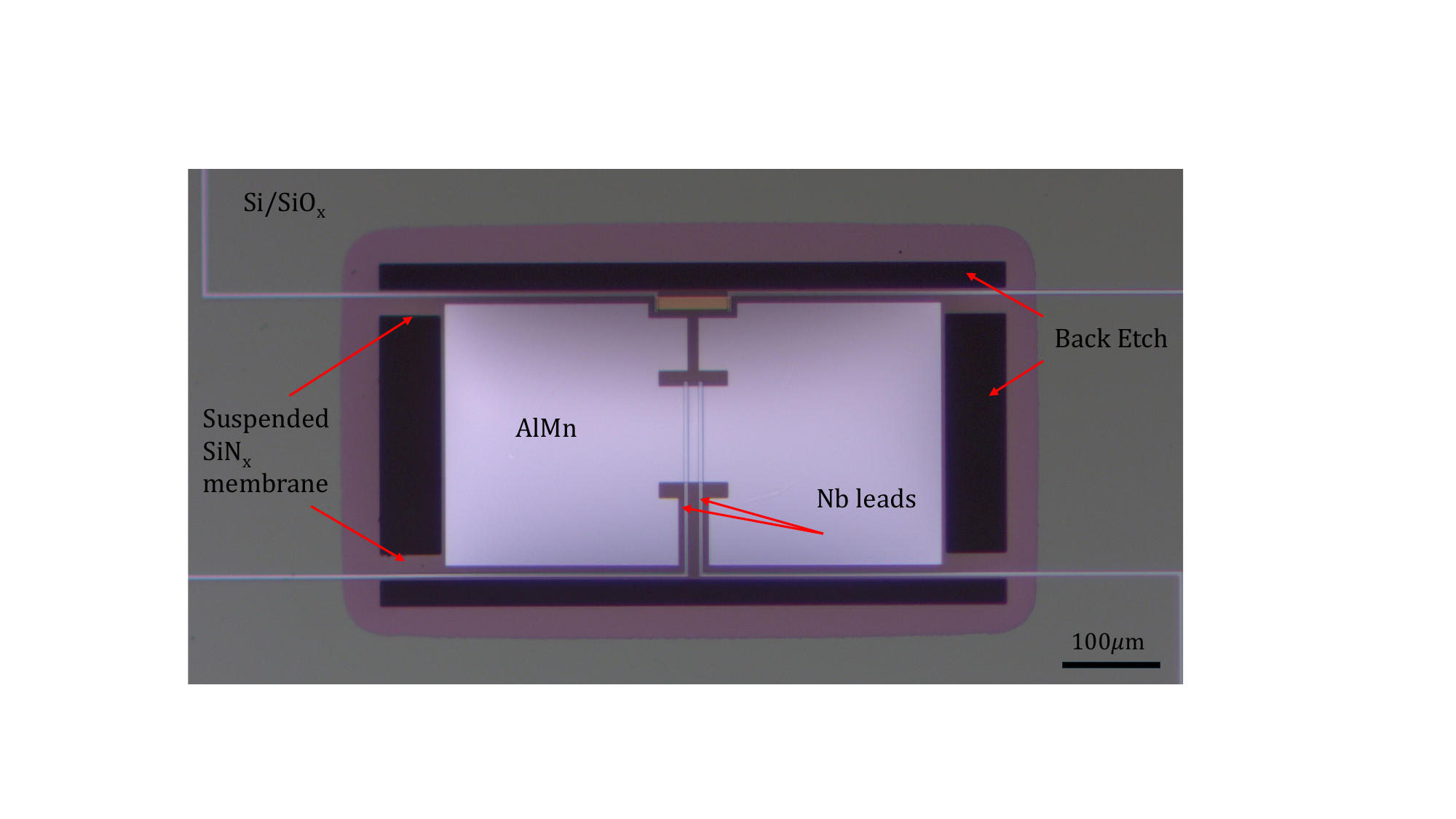}
    \caption{Optical image of the AlMn TES device studied in this work. This is a prototype device for millimeter-wavelength detection applications. The fabrication begins with Si wafers covered by \SI{450}{\nano\meter} of thermally grown SiO$_x$. \SI{2}{\micro\meter} of SiN$_x$ is deposited on top, followed by \SI{400}{\nano\meter} of AlMn with 2000 ppm (atomic ratio) Mn to Al, and \SI{200}{\nano\meter} of Nb. After etching, the resulting structure consists of a large $275\times\SI{500}{\micro\meter}$ plane of AlMn with $\SI{5}{\micro\meter}$ wide Nb leads spaced $\SI{10}{\micro\meter}$ apart, as measured from the inner edge of both leads. The Si wafer is then back etched, leaving the device structure suspended on a free-floating SiN$_x$ membrane \cite{Duff2016,Li2016}.}
    \label{fig:AlMn_optical_img}
\end{figure}

\begin{figure}[htbp]
    \includegraphics[width=.8\linewidth]{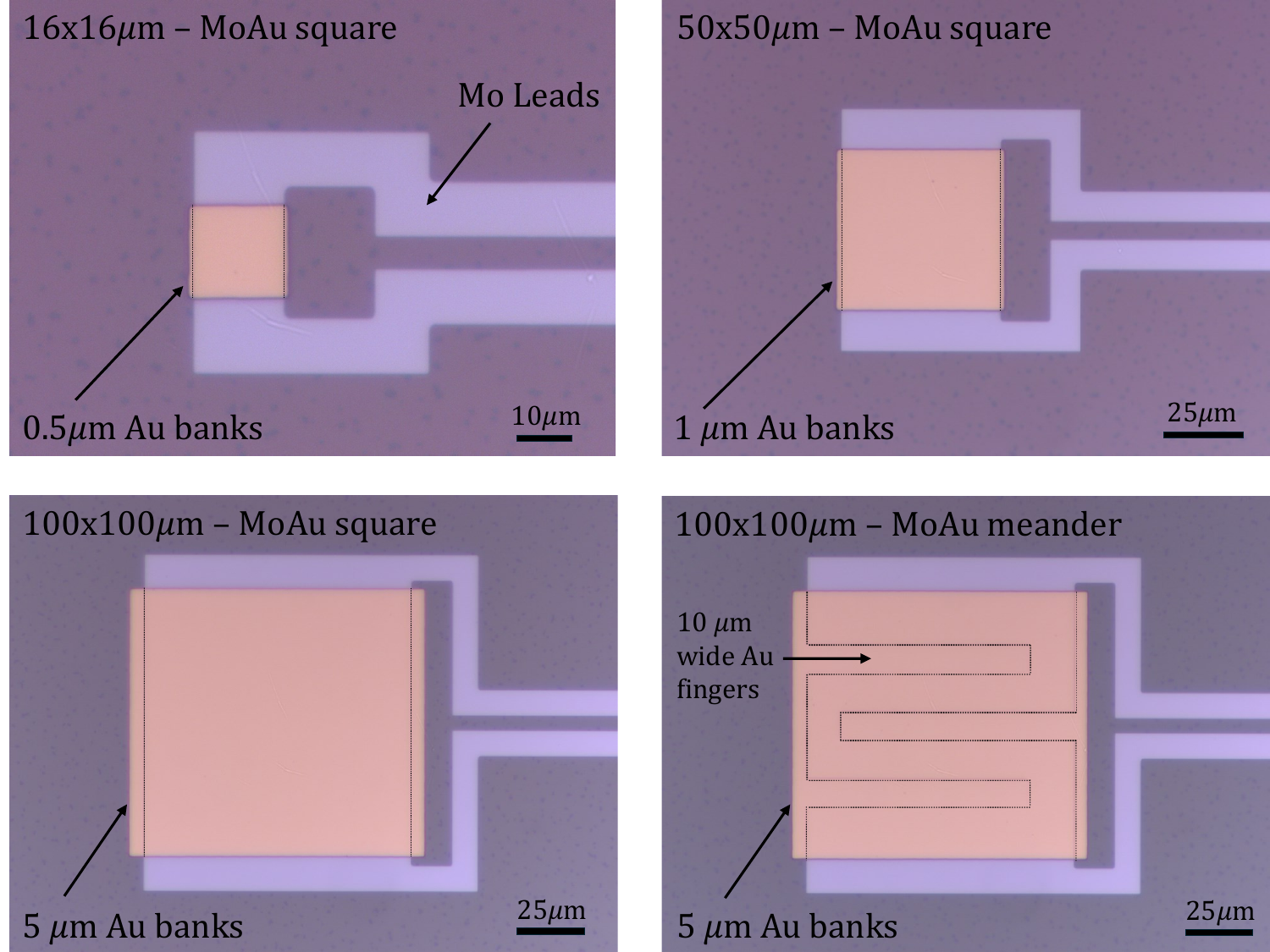}
    \caption{Optical images of the different MoAu TES device geometries studied in this work. These are prototype devices for x-ray detection applications. Here, \SI{46}{\nano\meter} of sputtered Mo is patterned into the shape of the leads and either a square or meandering shape in the TES device region. \SI{507}{\nano\meter} of Au was evaporated on top of the TES device region, forming a MoAu bilayer. The gold layer extends beyond the TES device region, forming Au only banks along the device edges with widths $\SI{0.5}{\micro\meter}$, $\SI{1}{\micro\meter}$, $\SI{5}{\micro\meter}$, $\SI{5}{\micro\meter}$ for the \SI{16}{\micro\meter} square, \SI{50}{\micro\meter} square, \SI{100}{\micro\meter} square, and \SI{100}{\micro\meter} meander TESs, respectively. In the \SI{100}{\micro\meter} meander device, the Au layer additionally lies in the cutouts of the underlying Mo meander, resulting in Au only fingers. The Si is back etched, suspending TES on a SiN$_x$ membrane \cite{Weber2020}.}
    \label{fig:all_optical_imgs}
\end{figure}

\clearpage
\newpage

\stepcounter{somecounter}
\section{Supplementary Note \thesomecounter: Scanning SQUID}\label{sec:SQUID}

\begin{figure}[htbp]
    \includegraphics[width=.33\linewidth]{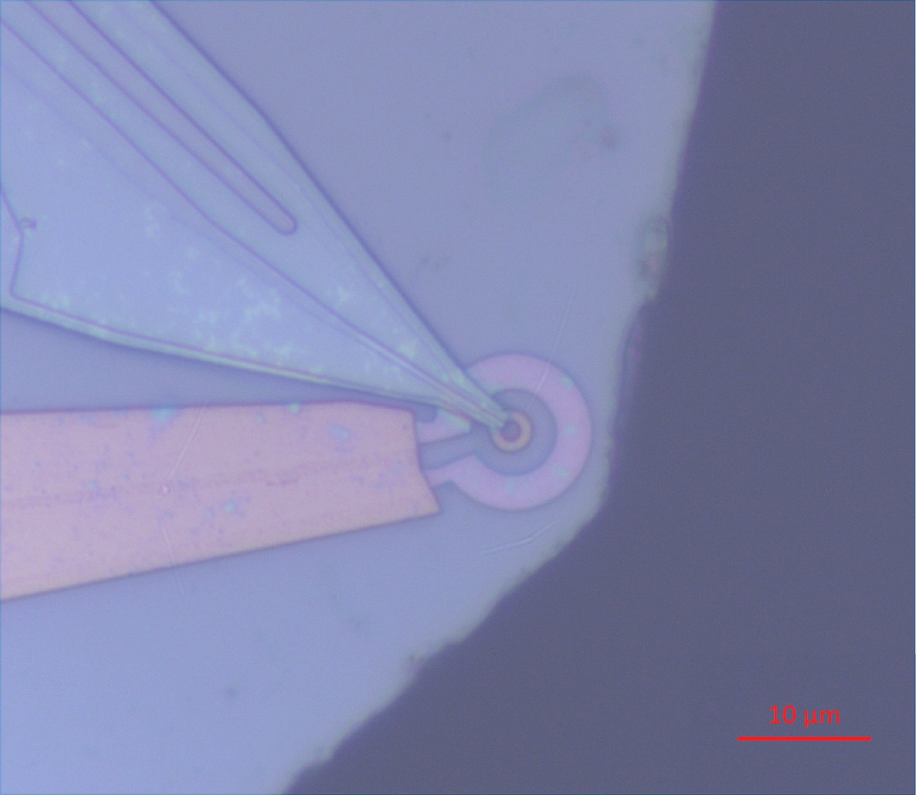}
    \caption{Optical image of the scanning SQUID susceptometer used in this work. The pickup loop has an inner and outer radius of $\SI{0.75}{\micro\meter}$ and $\SI{1.6}{\micro\meter}$ respectively; and the field coil has an inner and outer radius of $\SI{3.0}{\micro\meter}$ and $\SI{6.0}{\micro\meter}$ respectively. Scalebar is $\SI{10}{\micro\meter}$ }
    \label{fig:SQUID}
\end{figure}
 To measure the local diamagnetic response (DR), we apply an AC current through the field coil to apply a local field to the sample. The pickup loop, which is connected to a SQUID circuit, detects the resulting local magnetic field. For the data presented here, we used a current of \SI{86}{\micro\A} at \SI{277}{\hertz}. Far from the sample, we measure the bare mutual inductance $M_0$ between the field coil and the pickup loop by measuring the flux in the SQUID at the applied frequency using a lock-in amplifier.
 When the SQUID is brought near the surface of a superconductor, screening currents in the sample are induced over a length scale set by the field coil geometry. These screening currents reduce the local field coupled into the pickup loop, resulting in a reduction in the mutual inductance between the loops $\delta M$. In a sample with uniform superfluid stiffness, $\delta M$ can be modeled analytically \cite{kirtley2012hsweep} in terms of the sensor geometry and the superfluid stiffness (or equivalently penetration depth) of the sample. We discuss this modeling in more detail in Supplementary Section \ref{sec:Usadel}.C. 
 
 Near sample boundaries or in a spatially non-uniform superconductor, quantitatively interpreting the measured $\delta M$ becomes more complicated. In these cases, the screening currents in response to the field applied by the field coil flow through regions with varying superfluid stiffness in the superconductor or may be distorted due to a sample boundary. Although numerical modeling of the DR is possible in principle\cite{BISHOPVANHORN2022}, it requires detailed assumptions about the spatial profile of the superconducting properties and is therefore challenging. For the purpose of this work, the most significant consequence is a finite spatial resolution with which we can resolve changes in the superfluid density. The measured signal at one pixel location has contributions from screening currents flowing in nearby regions. For example, if the pickup loop is centered over a non-superconducting region (or a region with much weaker superfluid stiffness), the measured signal may be enhanced by adjacent regions with stronger screening.

\newpage
\newpage
\newpage

\stepcounter{somecounter}
\section{Supplementary Note \thesomecounter: Self-Consistent Usadel Modeling and Fitting to $\delta M(T)$ Data}\label{sec:Usadel}
\subsection{Modeling for AlMn Device using Usadel equations}
For the AlMn, we discuss modeling of the response of the AlMn device using Usadel equations. Given the device's aspect ration, we adopt an effective one-dimensional description. In the absence of magnetic fields and transport currents, so that the superconducting phase remains constant across the device, the Usadel equations reduce to a single differential equation for the function $\theta(\boldsymbol{x})$. This function describes the superconducting state and ranges from 0 to $\pi/2$, where $\theta=0$ corresponds to the normal state and $\theta=\pi/2$ to the fully developed superconducting state\cite{Kozorezov2011,MARTINIS200023,Harwin2017,Vodolazov2018,MARTINIS200023}. The resulting equation is given by:
\begin{equation}
    \frac{\hbar D}{2}\nabla^2\theta - \hbar\omega_n\sin{\theta} + \Delta\cos{\theta} = 0.
    \label{Usadel}
\end{equation}
Here, $D = D(\boldsymbol{x})$ is the local diffusion constant, $\Delta = \Delta(\boldsymbol{x})$ is the local superconducting gap, and $\omega_n = \frac{\pi k_B T}{\hbar}(2n+1)$ for $n\geq0$ and integer are the Matsubara frequencies. At the boundaries between the TES and the superconducting leads located at $\boldsymbol{x_0}$, $\theta(\boldsymbol{x})$ is subject to the boundary conditions \cite{Vodolazov2018}:
\begin{equation}
    D(\boldsymbol{x_0}-0)\frac{d\theta}{dx_{\boldsymbol{n}}}\bigg|_{\boldsymbol{x}=\boldsymbol{x_0}-0} = D(\boldsymbol{x_0}+0)\frac{d\theta}{dx_{\boldsymbol{n}}}\bigg|_{\boldsymbol{x}=\boldsymbol{x_0}+0}
\end{equation}
\begin{equation}
    \gamma_L \xi_{leads}\frac{d\theta}{dx_{\boldsymbol{n}}}\bigg|_{\boldsymbol{x}=\boldsymbol{x_0}\pm0} = \sin(\theta(\boldsymbol{x_0}+0)-  \theta(\boldsymbol{x_0}-0))
    \label{eqn:BC_leads}
\end{equation}
where $x_{\boldsymbol{n}}$ is a coordinate along the normal direction to the boundary, $\xi_{leads}$ is the coherence length in the leads, $\gamma_L$ characterizes the interface transparency between the TES and leads. The $\pm$ corresponds to the lead-TES and TES-lead interface respectively. On the external boundaries of the model geometry, we impose Neumann boundary conditions, $\frac{d\theta}{dx_{\boldsymbol{n}}}=0$.

Eq.\ref{Usadel} is solved along with the boundary conditions Eq.\ref{eqn:BC_leads} for each Matsubara frequency, giving a set of solutions $\frac{d\theta}{dx_{\boldsymbol{n}}}$. The gap $\Delta(\boldsymbol{x})$ is then calculated by:

\begin{equation}
    \Delta\ln{\frac{T}{T_c}} + 2\pi k_BT\sum_{\omega_n>0}(\frac{\Delta}{\hbar\omega_n}-\sin{\theta})=0
    \label{gap}
\end{equation}

Here, $T_c = T_c(\boldsymbol{x})$ denotes the local, nominal (non-proximitized) transition temperature of the different regions of the device, and $T$ is the sample temperature. The coupled equations Eq.\ref{Usadel} and \ref{gap} are solved self-consistently using a Newton's method algorithm for each temperature $T$, from which we obtain the temperature dependence of the gap at each spatial position $\Delta(T,\boldsymbol{x})$. The sum over Matsubara frequencies is truncated at $n=100$, and we verify that this cutoff is sufficient to obtain a convergence of the solution.

\subsection{Modeling for MoAu Device using Usadel equations}
For the square MoAu devices, we solve the coupled equations Eq.\ref{Usadel} and \ref{gap} in two dimensions. At the TES-lead interface, we impose the boundary conditions given in Eq.\ref{eqn:BC_leads}, while Neumann boundary conditions are applied at the external boundaries of the device. In the square MoAu devices, additionally normal metal banks are present. At the interfaces between the MoAu bilayer and the normal metal banks, weapply boundary conditions analogous to Eq.\ref{eqn:BC_leads}, given by:
\begin{equation}
    D(\boldsymbol{x_0}-0)\frac{d\theta}{dx_{\boldsymbol{n}}}\bigg|_{\boldsymbol{x}=\boldsymbol{x_0}-0} = D(\boldsymbol{x_0}+0)\frac{d\theta}{dx_{\boldsymbol{n}}}\bigg|_{\boldsymbol{x}=\boldsymbol{x_0}+0}
\end{equation}
\begin{equation}
    \gamma_B \xi_{TES} \frac{d\theta}{dx_{\boldsymbol{n}}}\bigg|_{\boldsymbol{x}=\boldsymbol{x_0}\pm0} = \sin(\theta(\boldsymbol{x_0}+0)-  \theta(\boldsymbol{x_0}-0))
    \label{eqn:BC_banks}
\end{equation}
where $\xi_{TES}$ is the coherence length in the TES, $\gamma_B$ describes the interface transparency between the TES and normal banks, and the $\pm$ corresponds to the TES-bank and bank-TES interface respectively. The system is then solved throughout the device, including in the banks, where the diffusion constant $D(\boldsymbol{x})=D_{banks}$ and the local critical temperature used in Eq.\ref{gap} is set very small, $T_{c_{banks}}=$\SI{0.1}{\milli\kelvin}.

\subsection{Model Parameters}
Eq.\ref{eqn:BC_leads} and \ref{eqn:BC_banks} describe a jump in $\theta$ that occurs at the S-S' and N-S' interface respectively. For simplicity, we set $\gamma_S$ and $\gamma_B$ to small values (both equal to 0.001), such that the interfaces are mostly transparent and $\theta$ is essentially continuous across them. Empirically, we find that significantly varying $\gamma_S$ and $\gamma_B$ plays only a small role in determining $\Delta$ at the center of the devices used to calculate the model $\delta M(T)$ curves in Fig.1(d) and Fig.3(b) in the main text. They do play a role in determining the gap closer to the interfaces, but quantitative modeling of the gap near the interfaces is beyond the scope of this work. We solve the device geometry in units of the coherence length of the leads, $\xi_{leads}^2 = \hbar D_{leads}/2\pi k_B T_{c_{leads}}$. Therefore, the value of $\xi_{leads}$, or equivalently $D_{leads}$, determines the effective device geometry. The relevant parameters that strongly influence the gap at the center of the device are therefore $\xi_{leads}$ and $D_{TES}$. For each of the MoAu TESs, we take $\xi_{leads} =$ \SI{0.5}{\micro\meter} and $D_{TES} = 1.4D_{leads}$. For the AlMn device, we take $\xi_{leads} =$\SI{50}{\nano\meter} and $D_{TES} = 1.5D_{leads}$.

\subsection{Calculating $\delta M$ from the Usadel gap $\Delta$}
For an isotropic BCS superconductor, the normalized superfluid density $\rho_s$ can be related to the superconducting gap and penetration depth $\lambda$ by \cite{Prozorov2006}:

\begin{equation}
    \rho_{s}(T) = \left({\frac{\lambda_0}{\lambda(T)}}\right)^{2} = 1 - \frac{1}{2k_B T} \int_{0}^{\infty}cosh^{-2}\left(\frac{\sqrt{\varepsilon^2+\Delta^2(T)}}{2k_BT}\right)\,d\varepsilon
    \label{eqn:rho(T)}
\end{equation}
where $\lambda_0$ is the zero temperature penetration depth.

Assuming a uniform in-plane penetration depth, the mutual inductance between the field coil and pickup loop of the SQUID as a function of SQUID-sample distance $z$ can be related to the penetration depth of the superconductor by \cite{kirtley2012hsweep}:

\begin{equation}
\delta M(z) = M_0 \int_{0}^{\infty} dx\,xe^{-2xz/a}J_1(x)\left[\frac{-(q+x)(q-x)+e^{2qt/a}(q-x)(q+x)}{-(q-x)(q-x)+e^{2qt/a}(q+x)(q+x)}\right]
\label{eq:integral}
\end{equation}

where $M_0$ is the bare mutual inductance between the field coil and pickup loop of the SQUID (398 $\phi_0/$A) measured far from the sample surface, $a$ is the effective field-coil radius, $t$ is the superconductor thickness, and $q$ is related to the penetration depth by $q = \sqrt{x^2+a^2/\lambda^2}$. At a fixed SQUID-sample distance $z=z^*$, we can then relate the measured $\delta M(T)$ to $\rho_s(T)$ through Eq.\ref{eqn:rho(T)} and \ref{eq:integral} with the gap $\Delta(T,\boldsymbol{x})$ obtained from solving the self-consistent Usadel equations.

\subsubsection{Results for $\delta M(T)$ for AlMn Device}

\begin{figure}[htbp]
    \includegraphics[width=1\linewidth]{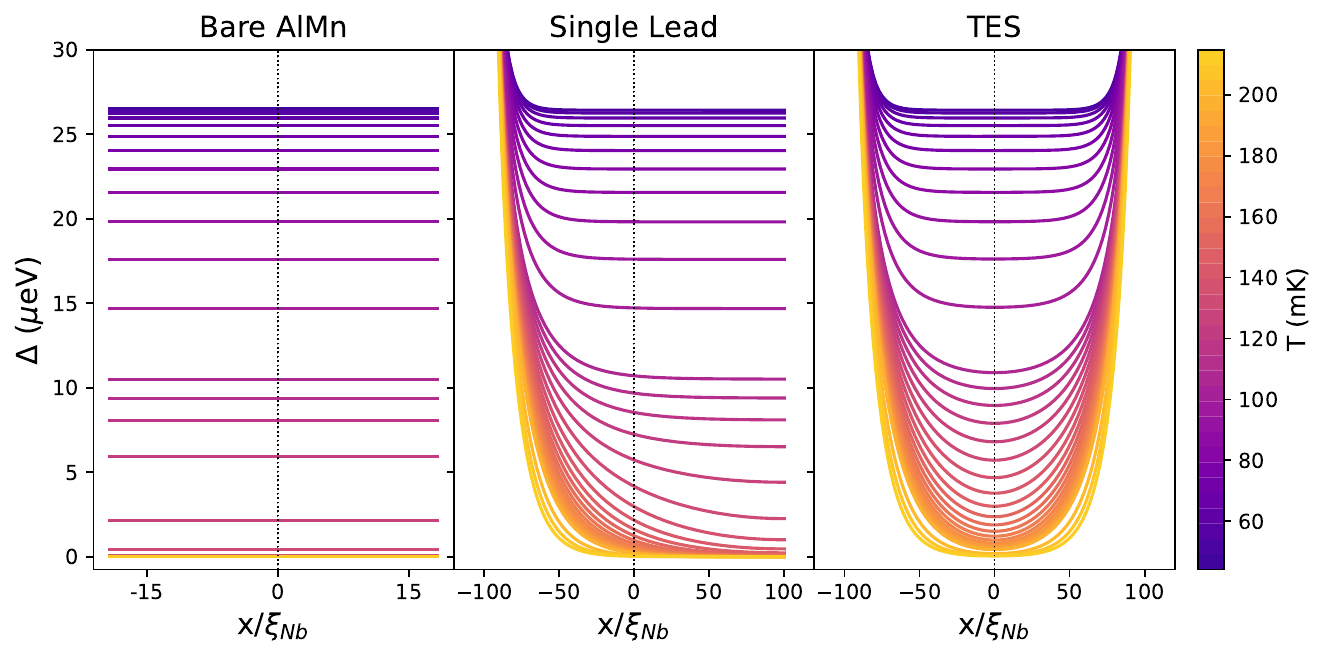}
    \caption{Superconducting gap $\Delta(x)$ calculated by solving the self-consistent Usadel equations for the bare AlMn with $T_{c_0} =$ \SI{174.5}{\m}K; an S-S' system where the AlMn sees the influence of a single Nb lead; and an S-S'-S system where the AlMn sees the influence of both Nb leads as in the TES device region. Taking $\Delta(x/\xi_{Nb} = 0)$ in these plots gives $\Delta(T)$ as plotted in Fig.1(e) in the main text. The temperatures used in the calculations correspond to the temperatures at which we acquired experimental data.}
    \label{fig:1D_usadel_gap}
\end{figure}

\begin{figure}
    \includegraphics[width=1\linewidth]{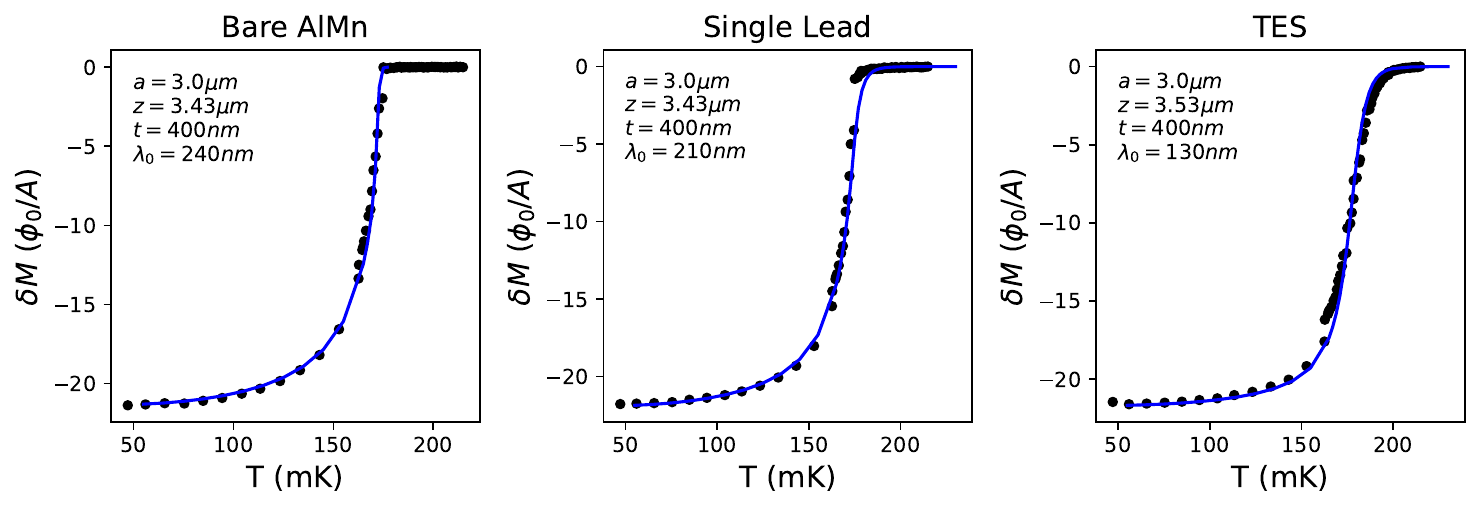}
    \caption{Model fits for the $\delta M(T)$ data based on the superconducting gap calculated using Usadel equations similar to Fig.1(d) in the main text. In Fig.1(d) $\delta M(T)$ is normalized to its value measured at \SI{55}{\milli\kelvin}, whereas here we show the unnormalized $\delta M(T)$. Fit parameters are indicated in the insets. Note that a different $\lambda_0$ is used in each case as discussed below and in the main text.}
    \label{fig:1D_deltaM}
\end{figure}

We use the gap shown in Fig.1(e) of the main text (See Fig.\ref{fig:1D_usadel_gap}) together with Eq.\ref{eqn:rho(T)} and \ref{eq:integral} to generate model curves for $\delta M(T)$ shown in Fig.\ref{fig:1D_deltaM}. The parameters entering the $\delta M(T)$ model curve are the field-coil radius $a$, which we fix to the inner radius of the the field-coil geometry; the SQUID-sample distance $z$, which is allowed to vary; the sample thickness $t$ which we set to the full AlMn thickness of \SI{400}{\nano\meter}; and the zero temperature penetration depth $\lambda_0$, which is also allowed to vary. 
With these assumptions, the fitted value for $z$ varies only by $\sim$\SI{100}{\nano\meter}, consistent with a slight misalignment in the scan plane which would introduce slight height differences along the 1D line cuts. In addition, we suspect some warping of the SiN$_x$ membrane, as suggested by the slight waviness in the 1D line cuts (e.g. see low temperature line cuts in Fig.1(c) of the main text), which cannot be accounted by a simple plane tilt. 

In the Usadel modeling of the gap (Fig.1(e)), the $\Delta(T)$ for the bare AlMn, the single lead proximity case, and the full TES geometry converge to the same curve below $\sim$\SI{150}{\milli\kelvin}. In contrast, the measured $\delta M(T)$ curves exhibit different curvatures at low temperatures, most clearly visible in Fig.1(d). The discrepancy  can be reproduced in the model by assuming different values for the zero-temperature penetration depth $\lambda_0$ for the three cases. However, such variation in $\lambda_0$ is not obviously physically motivated.

\begin{figure}
    \includegraphics[width=1\linewidth]{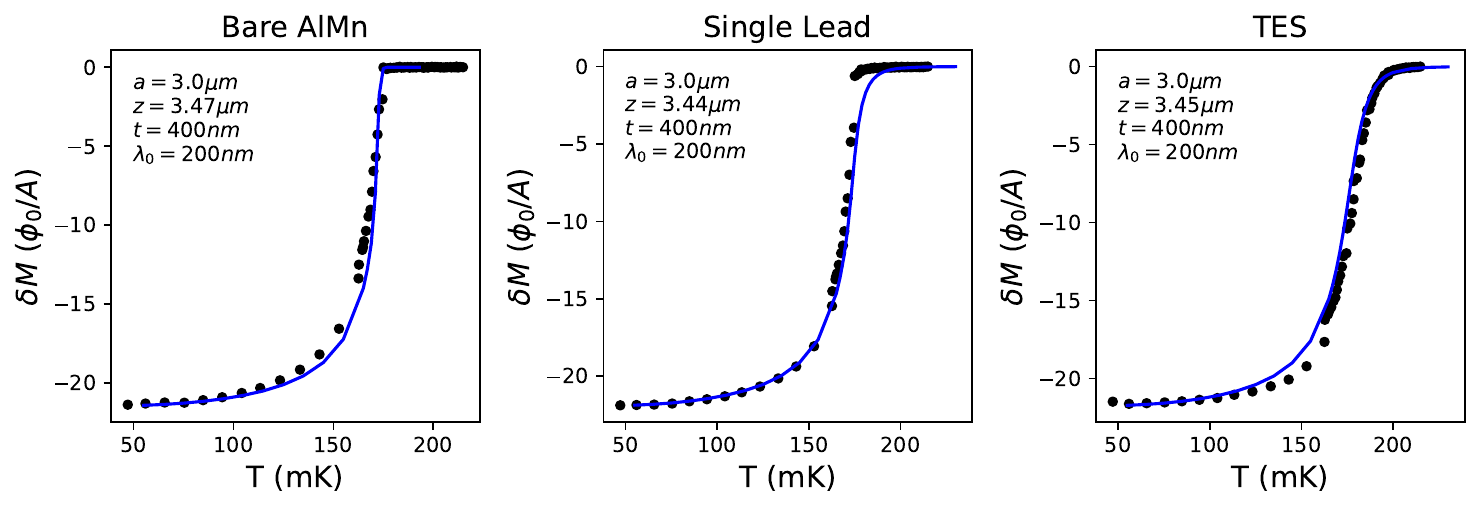}
    \caption{Model $\delta M(T)$ curves similar to Fig. \ref{fig:1D_deltaM}, now calculated using  a spatial average of the gap $\Delta(x)$ over \SI{5}{\micro\meter}, as obtained from the Usadel euqations. We find reasonable agreement with the data while using the same value of $\lambda_0$. All fit parameters are indicated in the insets.
    }
    \label{fig:1D_deltaM_mean}
\end{figure}

Another possibility is that this discrepancy arises from the finite size of the field coil and pickup loop of the SQUID such that our measurement includes an effective spatial average over the sample and contains contributions from more proximitized regions closer to the leads. In Fig.\ref{fig:1D_deltaM_mean}, we take the mean of the gap $\Delta(x)$ at each temperature over a \SI{5}{\micro\meter} radius as a simple way to include some amount of spatial averaging and use it to calculate the model $\delta M(T)$ curves. Now, using the same $\lambda_0$ for each case, we find reasonable agreement between the data across the different geometries. We note that a precise model of $\delta M(T)$ that accounts for the finite size of the SQUID susceptometer and the spatial inhomogeneity in a comparably small device is difficult, but this simple method qualitatively explains some detailed aspects of the data. 

\FloatBarrier
\clearpage
\subsubsection{Results for $\delta M(T)$ for MoAu Square Device}
\begin{figure}[htbp]
    \includegraphics[width=1\linewidth]{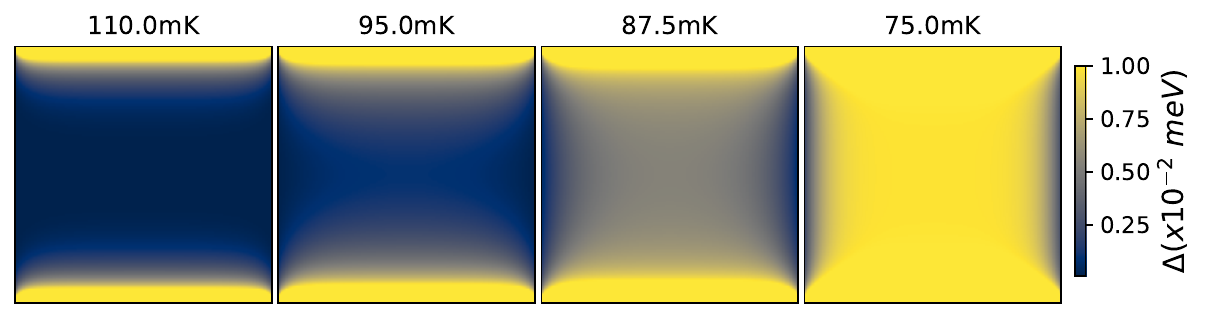}
    \caption{Calculated superconducting gap $\Delta(\boldsymbol{x})$ obtained from solving the self-consistent Usadel equations in two dimensions for the 50x50$\mu$m square device at selected temperatures. We extract the value of $\Delta(T)$ at the center of the device, and use it, together with values for additional temperatures (Fig.3(b)), to calculate the model $\delta M(T)$ shown in Fig.3(a) of the main text. The gap exhibits a hourglass feature similar to what we observe in the imaged diamagnetic response of the device.}
    \label{fig:usadel_gap}
\end{figure}

\begin{figure}[htbp]
    \includegraphics[width=.3\linewidth]{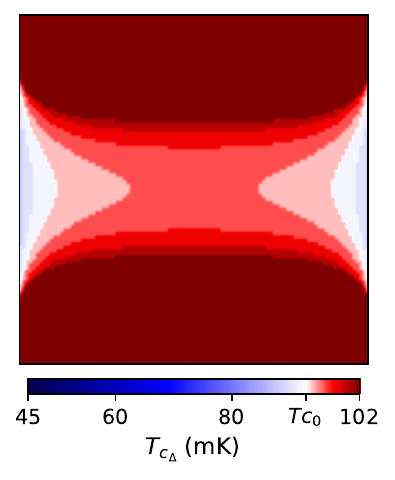}
    \caption{Map of the highest temperature, $T_{c_\Delta}$, at which the gap calculated from the Usadel model $\Delta(x,y)$ (see Fig.\ref{fig:usadel_gap}) exceeds a threshold value of $\Delta =$\SI{0.5}{\micro\electronvolt}. This map is analogous to the $T_{c_{GL}}$ maps. However, because solving the Usadel equations in two dimensions is significantly more computationally demanding than the GL modeling, the temperature spacing and range are more limited. As a result, Fig.\ref{fig:usadel_Tc} contains a limited number of temperature contours. The $T_{c_{\Delta}}$ map exhibitis similar features to the $T_{c_{GL}}$ maps.}
    \label{fig:usadel_Tc}
\end{figure}
\newpage



\begin{figure}[htbp]
    \includegraphics[width=0.8\linewidth]{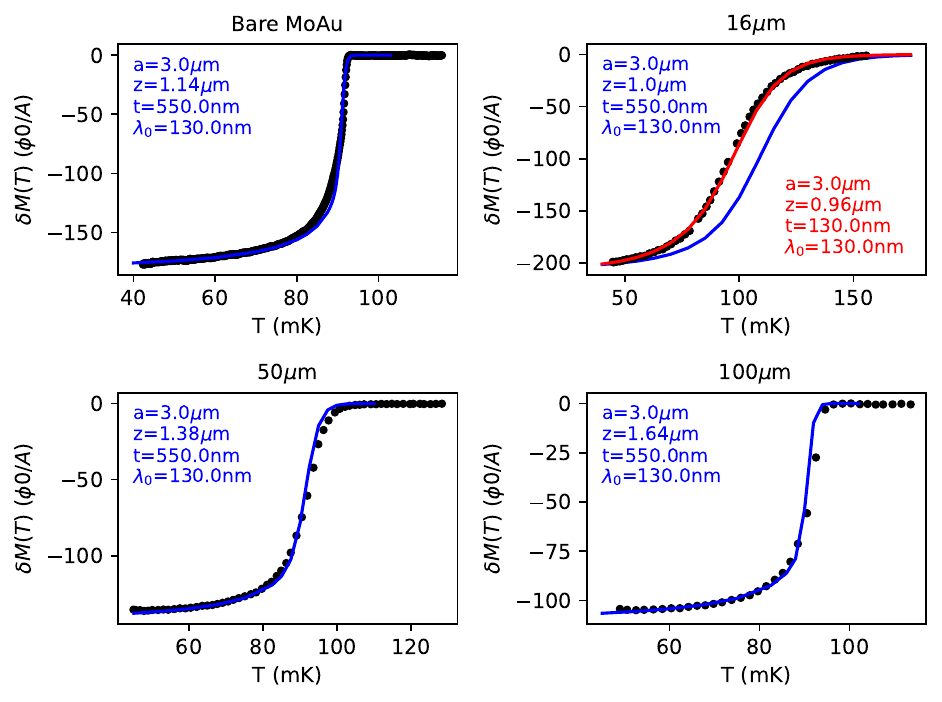}
    \caption{Model fits for the $\delta M(T)$ data as shown in Fig.3(a) in the main text. In Fig.3(a) $\delta M(T)$ is normalized to its value measured at \SI{50}{\milli\kelvin}, whereas here we show the unnormalized $\delta M(T)$. . Assuming the same sample thickness and $\lambda_0$ for all samples (blue curves) yields good agreement with all but the \SI{16}{\micro\meter} device. Assuming the same $\lambda_0$, but a different effective sample thickness for the \SI{16}{\micro\meter} device (red curve), results in good agreement with all data, suggestive of a temperature dependence of the superconductivity through the bilayer, as discussed in the text.}
    \label{fig:fit_share_all_params}
\end{figure}


In Fig.3(a) of the main text, we plot the temperature dependent DR $\delta M(T)$ measured at the center of the three square MoAu bilayer TES geometries (\SI{16}{\micro\meter}, \SI{50}{\micro\meter}, and \SI{100}{\micro\meter}) along with that measured on the bare MoAu film without any other nearby superconducting or normal metal structures. We model the proximity effects in these geometries by solving the self-consistent Usadel equations in 2D, including both the influence of the superconducting leads and the normal banks. From the Usadel model, we obtain a spatially varying superconducting gap $\Delta(\boldsymbol{x})$ for each temperature (Fig.\ref{fig:usadel_gap}). Taking the central pixel of the square grid we obtain $\Delta(T)$ for each device (Fig.3(b)). We then model our $\delta M(T)$ data using Eq.\ref{eqn:rho(T)} and \ref{eq:integral}. We can model the bare MoAu, \SI{50}{\micro\meter}, and \SI{100}{\micro\meter} devices well assuming the same penetration depth $\lambda_0$ and that the sample thickness is that of the full MoAu bilayer. Using these same parameters for the \SI{16}{\micro\meter} yields a model $\delta M(T)$ curve with a similar general shape as the data, but with quantitative differences. 

All measured devices are fabricated on the same chip from the same MoAu bilayer. It is therefore reasonable to assume the same effective thickness of the superconducting layer, $t$, for all  devices. However, the MoAu bilayer itself is a proximity system, and the spatial distribution of superconductivity through the thickness of the MoAu layer and its evolution with temperature are not known. At low temperature, we expect the entire MoAu layer to be superconducting in all devices. This is supported by the observation that the \SI{10}{\micro\meter} wide Au fingers in the meander geometry become superconducting. Closer to and above $T_{c_0}$, however, the situation may differ. It is plausible that superconductivity is more confined to the Mo layer at elevated temperatures, with the Au layer  becoming only gradually proximitized upon cooling. In that case, the effective thickness of the superconducting layer is smaller near the transition and increases with decreasing temperature. 

Capturing such behavior would likely require a full 3D Usadel calculation to include spatial structure throughout the bilayer thickness.
Introducing a temperature-dependent effective thickness into a simplified model would add additional, poorly constrained parameters and naturally improve the fit quality. For this reason, we choose to model the devices using a fixed thickness.

However, for the smallest device, this effect is likely more pronounced because the DR emerges at temperatures significantly higher than $T_{c_0}$.  Allowing the \SI{16}{\micro\meter} device to have a reduced effective thickness of $t=$\SI{130}{\nano\meter}, while keeping the same value of $\lambda_0$ that successfully describes the other devices, yields good agreement with the data as shown by the red curve in Fig.\ref{fig:fit_share_all_params}. This issuggestive of a reduced effective thickness of the superconducting layer at higher temperatures in the smallest device.

\newpage
\clearpage

\newpage
\stepcounter{somecounter}
\section{Supplementary Note \thesomecounter: Ginzburg-Landau Modeling of superconducting transition in MoAu TES devices}\label{sec:GL}
We model the spatial variations in the superconducting transition of the MoAu TES devices using the Ginzburg-Landau (GL) equations. In the absence of magnetic fields and currents, such that the superconducting phase is constant, the GL equations reduce to a single nonlinear differential equation \cite{Tinkham}:

\begin{equation}
    \xi^2(T)\nabla^2\psi \pm \psi - \psi^3 = 0.
\label{eq:GL} 
\end{equation}

Here, $\psi$ is the superconducting order parameter normalized to its bulk value, $\xi=\xi(\boldsymbol{x})$ is the local coherence length, and $\boldsymbol{x}$ is the 2D coordinate within the device. We solve Eq.\ref{eq:GL} throughout all regions of the device including the MoAu bilayer, the Au banks, and Au fingers. The $\pm$ sign differentiates between two regimes: the $-$ sign applies where the underlying material is nominally normal, and the $+$ sign applies in regions where it is nominally superconducting. In the MoAu bilayer, this means that the $-$ sign applies for $T>T_{c_0}$, and the $+$ sign for $T<T_{c_0}$, where $T_{c_0}$ is the intrinsic critical temperature of the MoAu bilayer. For the Au banks and fingers, we take $T_{c_{0_{Au}}}=0$, such that these regions are treated as nominally normal at all temperatures. Therefore, we solve the equation using the $-$ sign in those regions for all temperatures.

Near $T_{c_0}$, we assume that the temperature dependence of the coherence length of the superconducting bilayer follows $\xi_{MoAu}(T) = \zeta_{MoAu}/\sqrt{|T-T_{c_0}|}$. For the coherence length of the normal Au regions, we assume the dirty limit and use $\xi_{Au} = \zeta_{{Au}}/\sqrt{T}$. Note that $\zeta_{MoAu}$ and $\zeta_{Au}$ are proportionality constants, not coherence lengths and have units \SI{}{\micro\meter}$\sqrt{K}$. At the interfaces between the MoAu bilayer and the Mo leads, we impose Dirichlet boundary conditions, which remain superconducting at all temperatures of interest, by setting $\psi = \psi_0 = 1$. At the interfaces between the MoAu bilayer and the Au regions, we employ de Gennes boundary conditions \cite{de_Gennes}:

\begin{equation}
    \frac{d\psi}{dx_{\boldsymbol{n}}}\bigg|_{\boldsymbol{x_0}} = \pm\frac{\psi(\boldsymbol{x_0})}{b},
\label{eq:de Gennes}
\end{equation}
where $x_{\boldsymbol{n}}$ is the direction normal to the interface, $\boldsymbol{x_0}$ denotes the position of the interface, and $b$ is the de Gennes extrapolation length. The sign $\pm$ is determined by the slope expected for $\psi$ at the interface, i.e. a positive slope at a N-S interface, where the order parameter grows moving across the interface into the superconducting region, and a negative slope for a S-N interface, where the order parameter decreases moving across the interface into the normal region. The extrapolation length $b$ characterizes the slope of $\psi$ at the interface and is a measure of the interface transparency. On the MoAu side, $b = b_{Au}$, the extrapolation length into the Au region, and on the Au side, $b = b_{MoAu}$, the extrapolation length into the MoAu region. At the interface between the device and the vacuum, we impose a Neumann boundary condition $\frac{d\psi}{dx_{\boldsymbol{n}}}= 0$.

We solve Eq.\ref{eq:GL} for the same set of temperatures used in the susceptibility imaging shown in Fig.\ref{fig:50um_suscep_imgs} and Fig.\ref{100um_meander_suscep_imgs}. At each position, we determine the temperature at which $\psi$ exceeds a threshold value $\epsilon$ to compose a $T_{c_{GL}}$ map. We compare this map to the measured $T_{c_{\chi}}$ map as discussed in the main text. 

The calculated $T_{c_{GL}}$ map depends on five input parameters: $\zeta_{MoAu}$, $\zeta_{Au}$, $b_{MoAu}$, $b_{Au}$, and $\epsilon$. We use the same values for these parameters to generate the $T_{c_{GL}}$ maps for the \SI{50}{\micro\meter} square device (Fig.2(c)) and the \SI{100}{\micro\meter} meander device (Fig.4(c)) in the main text: $\zeta_{MoAu} =$ \SI{0.6}{\micro\meter\sqrt{K}}, $\zeta_{Au} =$ \SI{1.0}{\micro\meter\sqrt{K}}, $b_{MoAu} =$ \SI{3}{\micro\meter}, $b_{Au} =$ \SI{1}{\micro\meter}, $\epsilon =$ 0.065.

We note that  $\zeta_{Au}$ and $b_{MoAu}$ primarily govern the behavior of the order parameter in the Au regions of the device (banks and fingers) and play only a minor role in the MoAu regions. Consequently, the $T_{c_{GL}}$ contours inside the MoAu square or meander are predominantly determined by $\zeta_{MoAu}$, $b_{Au}$, and $\epsilon$, whereas contours inside the Au banks and fingers are determined almost entirely by $\zeta_{Au}$, $b_{MoAu}$, and $\epsilon$. In practice, this means that the GL solutions in the MoAu regions  depend effectively only on three parameters.

The parameters $\zeta_{MoAu}$ and $\epsilon$ are strongly correlated: choosing a higher (lower) threshold for $\epsilon$ requires a correspondingly larger (smaller) value of $\zeta_{MoAu}$ to reach the threshold value of $\psi$ at the temperature observed in the data. Therefore, the value  $\zeta_{MoAu} =$ \SI{0.6}{\micro\meter\sqrt{K}} used here should not be interpreted as a quantitative estimate of the actual coherence length of the MoAu bilayer, as its value is dependent on the choice for $\epsilon$. 

\begin{figure}[htbp]
    \includegraphics[width=1\linewidth]{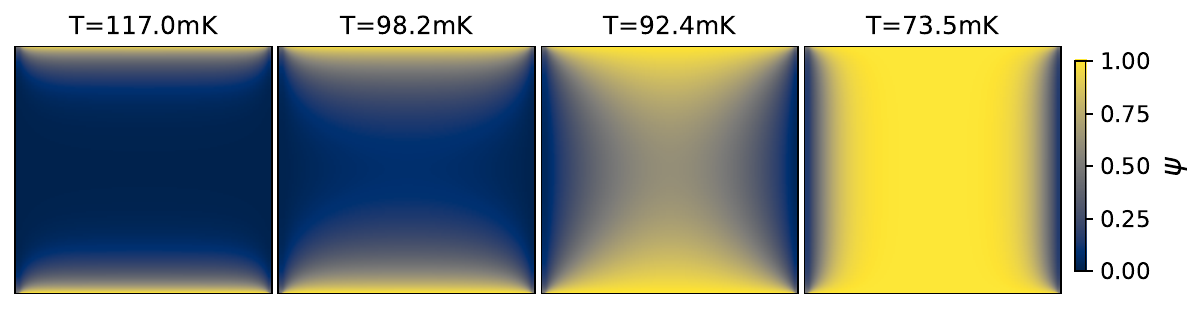}
    \caption{Calculated GL order parameter $\psi$ within the \SI{50}{\micro\meter} square device at selected temperatures. The temperatures correspond to those at which images of the DR $\delta M$ are presented in main text. We apply a threshold to $\psi$ to identify regions of the sample that become superconducting at each temperature, from which the $T_{c_{GL}}$ map is constructed.}
    \label{Fig:50_GL_psi}
\end{figure}

\begin{figure}[htbp]
    \includegraphics[width=1\linewidth]{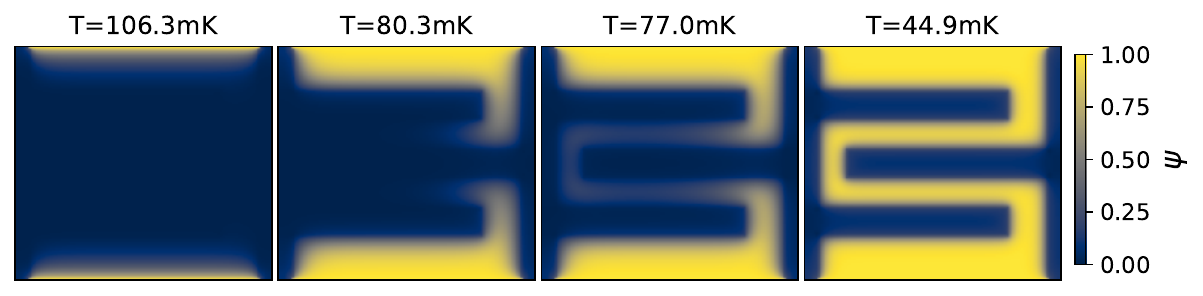}
    \caption{Same as Fig.\ref{Fig:50_GL_psi} but for the meander device.}
    \label{Fig:meander_GL_psi}
\end{figure}

\newpage

\newpage
\stepcounter{somecounter}
\section{Supplementary Note \thesomecounter: Imaging in other device geometries}\label{sec:other_devices}

\begin{figure}[htbp]
    \includegraphics[width=0.6\linewidth]{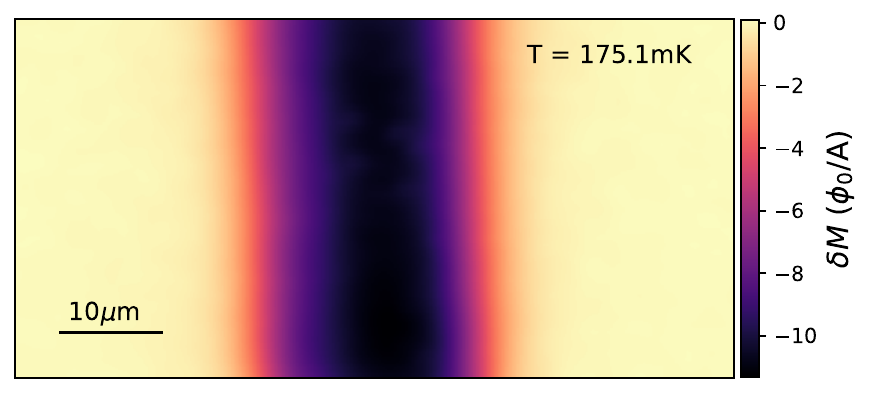}
    \caption{Image of the diamagnetic response of the AlMn TES described in the main text. At the shown temperature, \SI{175.1}{\milli\kelvin}, the region in between the Nb leads is proximitzed. Near the device center shown here, the response is translationally invariant response along the width (long dimension, vertical here). This observation motivates treating the system as effectively one-dimensional in the Usadel modeling and using one-dimensional line cuts to characterize the device in the main text.}
    \label{Fig:AlMn_img}
\end{figure}

\begin{figure}[htbp]
    \includegraphics[width=1\linewidth]{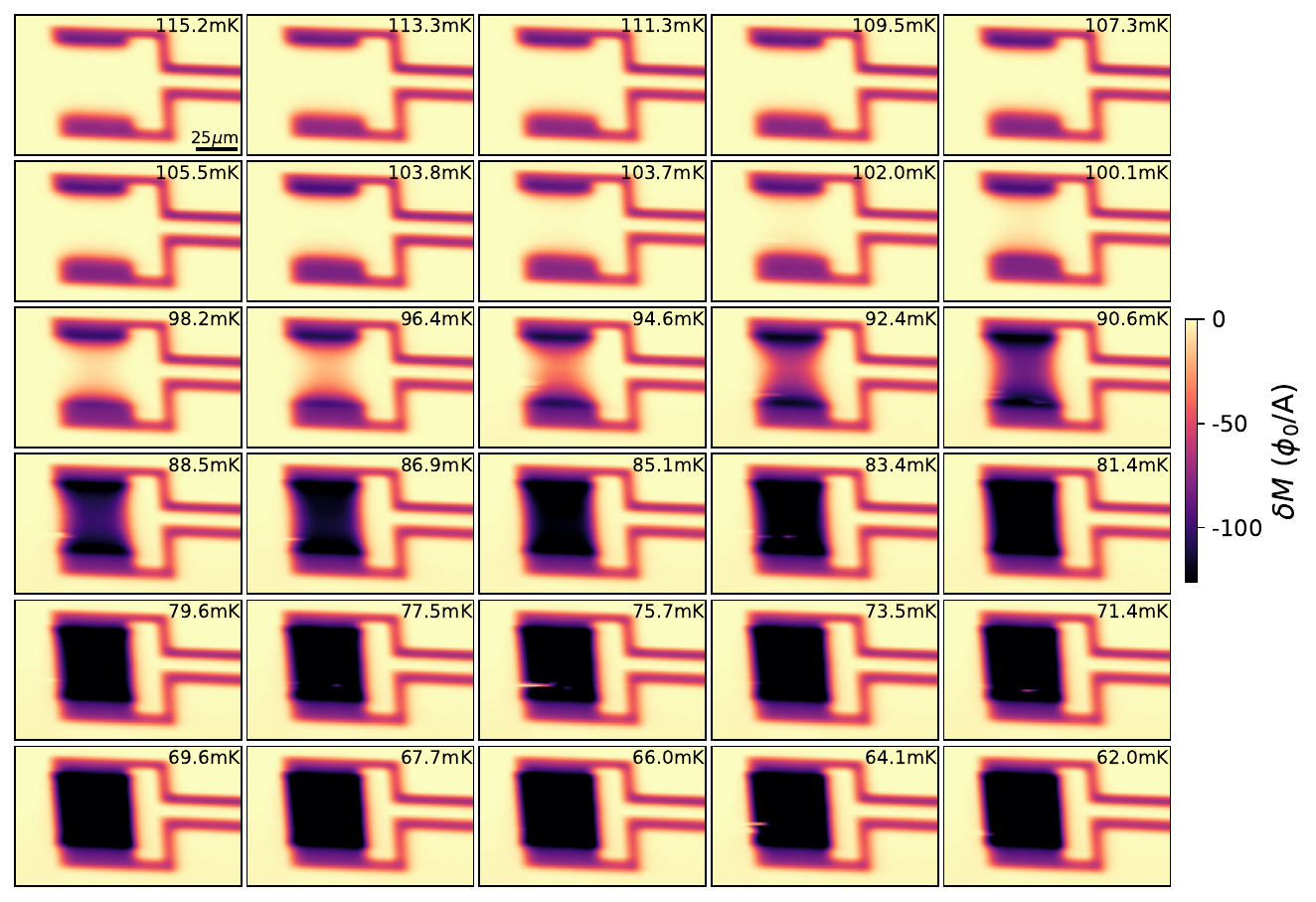}
    \caption{Full image series of the diamagnetic response of a \SI{50}{\micro\meter} MoAu square TES device described in the main text. This dataset is used to generate the $T_{c_{\chi}}$ map. The normal metal banks have a width of \SI{1}{\micro\meter}.}
    \label{fig:50um_suscep_imgs}
\end{figure}

\begin{figure}[htbp]
    \includegraphics[width=1\linewidth]{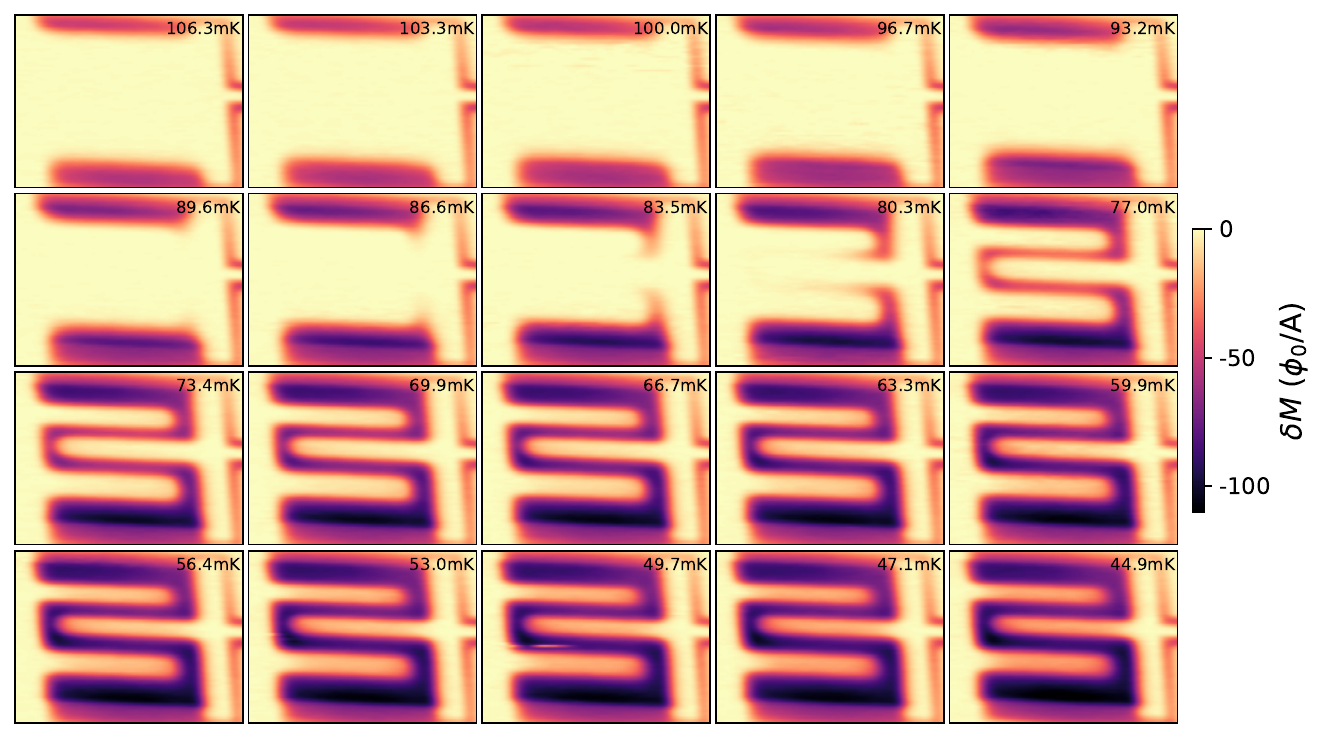}
    \caption{Image series of the diamagnetic response of the \SI{100}{\micro\meter} MoAu meander TES device as described in the main text. This dataset is used to generate the $T_{c_{\chi}}$ map. The normal metal banks have a width of \SI{5}{\micro\meter}.}
    \label{100um_meander_suscep_imgs}
\end{figure}

\begin{figure}[htbp]
    \includegraphics[width=.8\linewidth]{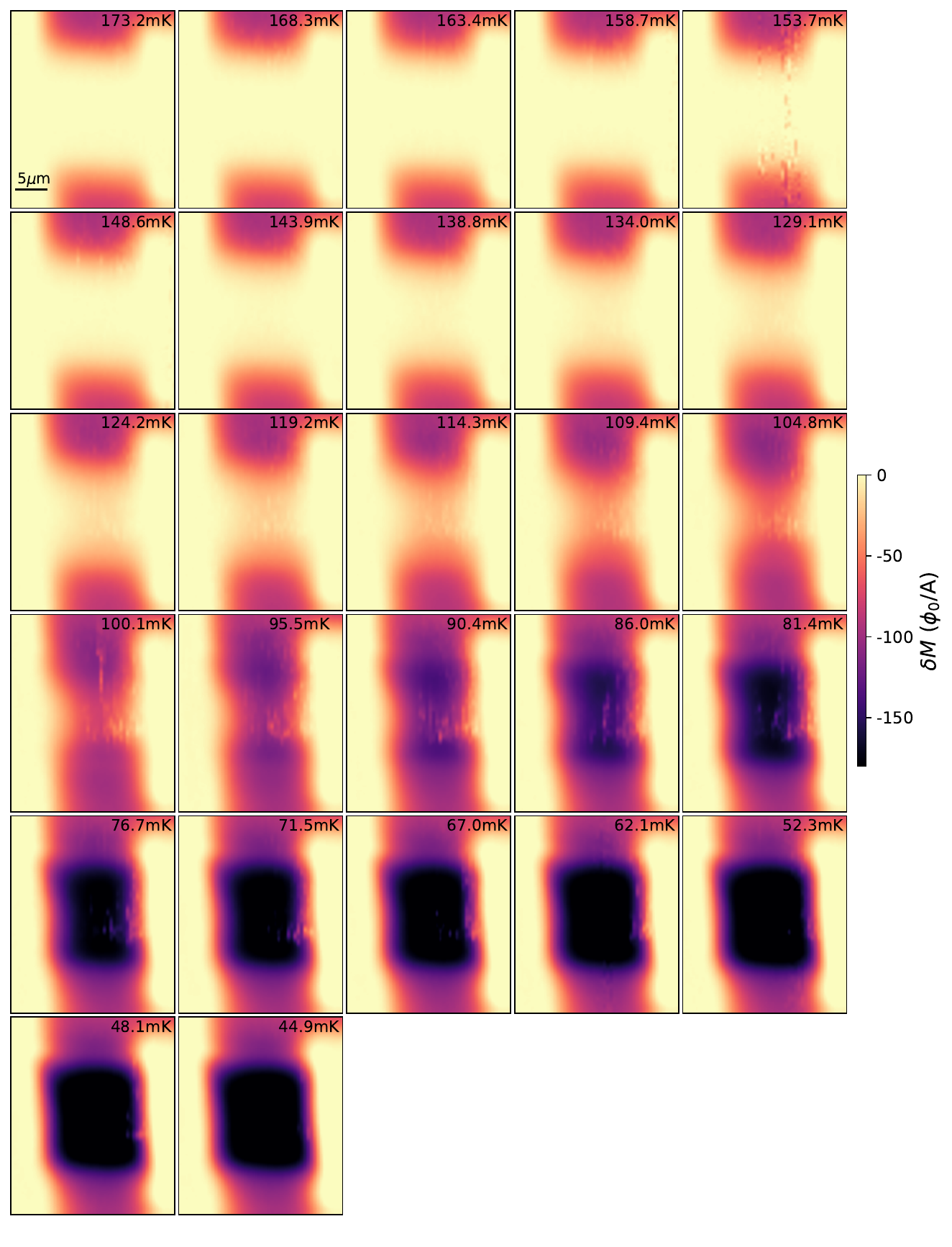}
    \caption{Image series of the diamagnetic response of a MoAu \SI{16}{\micro\meter} square TES device. The width of the normal metal banks is \SI{0.5}{\micro\meter}. We observe similar qualitative behavior as for the \SI{50}{\micro\meter} device described in the main text, but over a larger temperature range.}
    \label{fig:16um_suscep_imgs}
\end{figure}

\begin{figure}[htbp]
    \includegraphics[width=1\linewidth]{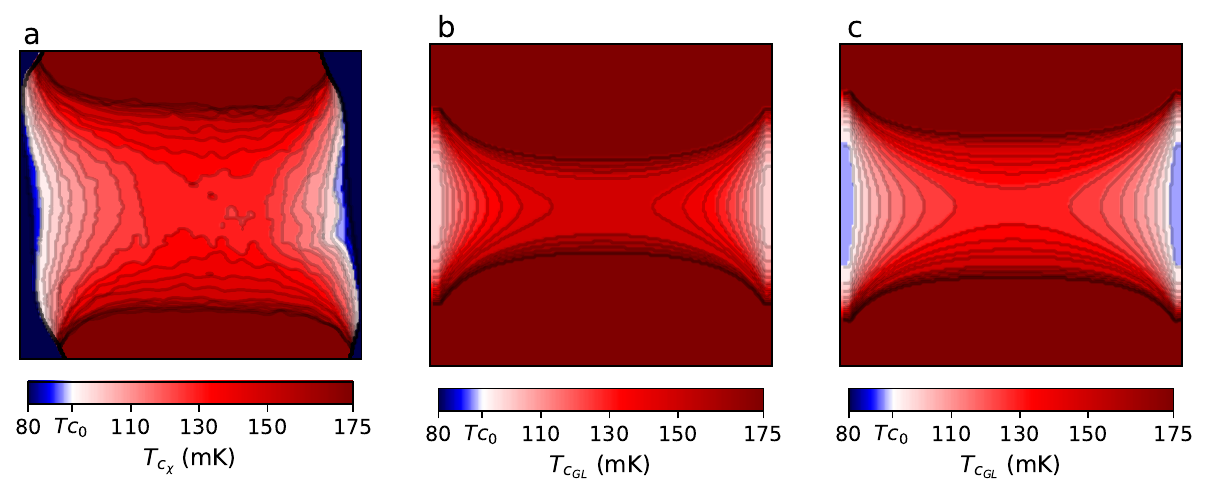}
    \caption{$T_{c_{\chi}}$ map of the 16x16$\mu$m square device obtained from the image series in Fig.\ref{fig:16um_suscep_imgs}. A similar 'hourglass' feature of the local $T_{c_{\chi}}$ is observed from the combined proximity and inverse proximity effects from the leads and banks respectively. However, most of the device exhibits $T_{c_{\chi}} > T_{c_0}$ and therefore appears red on the color scale (white denotes $T_{c_0}$). This suggests a more enhanced $T_{c_{\chi}}$ from the proximity effect from the leads in the smaller device. (b) $T_{c_{GL}}$ map calculated by solving the GL equations as described in Supp. Sec.\ref{sec:GL}, using the same GL parameters as used for the $T_{c_{GL}}$ maps in the main text. The resulting $T_{c_{GL}}$ reproduces some features of the measured $T_{c_{\chi}}$, but does not fully capture the data. This discrepancy may be a product of the elevated transition temperatures in this device, placing the GL analysis further away from $T_{c_0}$ where its validity and/or the assumed temperature dependence of $\xi(T)$ begins to break down. (c) Adjusting the threshold parameter to $\epsilon=0.11$ yields a $T_{c_{GL}}$ map that more closely resembles the measured $T_{c_{\chi}}$ map in (a). A similar $T_{c_{GL}}$ map can be obtained by keeping $\epsilon=.065$ and instead changing only $\zeta_{_{MoAu}}$ (See Supp. Sec.\ref{sec:GL} for details).}
    \label{fig:16um_Tc_map}
\end{figure}

\begin{figure}[htbp]
    \includegraphics[width=1\linewidth]{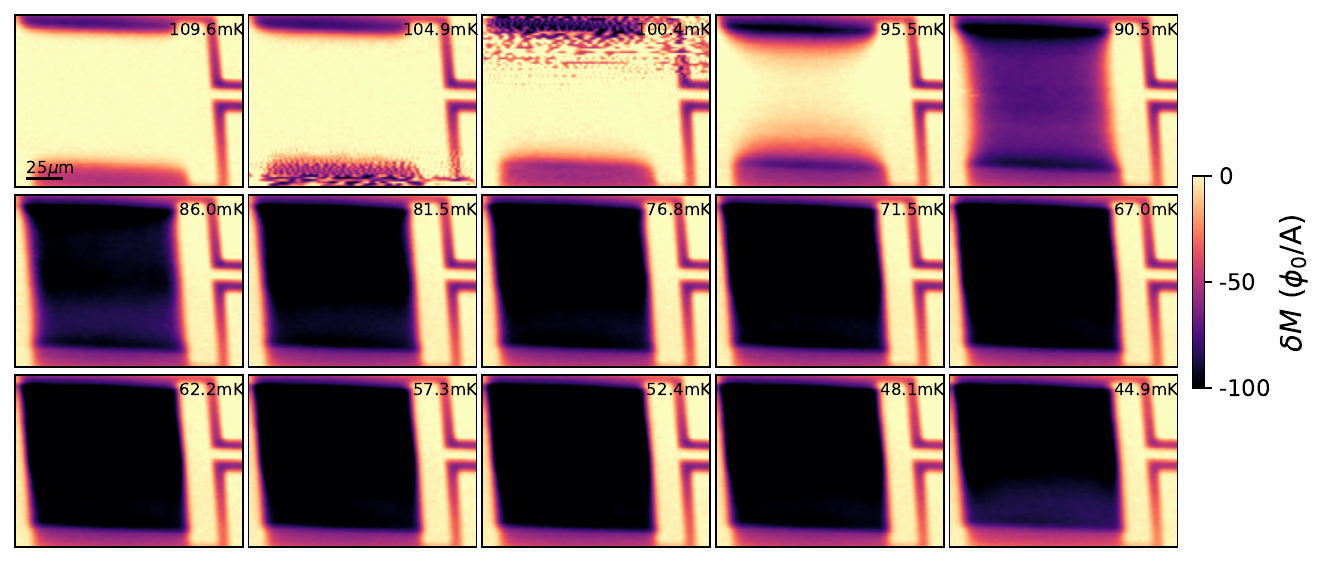}
    \caption{Image series of the diamagnetic response of a MoAu \SI{100}{\micro\meter} square TES device. The width of the normal metal banks is \SI{5}{\micro\meter}. We observe similar qualitative behavior as in the \SI{50}{\micro\meter} device described in the main text, but over a smaller range of temperatures.}
    \label{fig:100um_suscep_imgs}
\end{figure}

\begin{figure}[htbp]
    \includegraphics[width=0.8\linewidth]{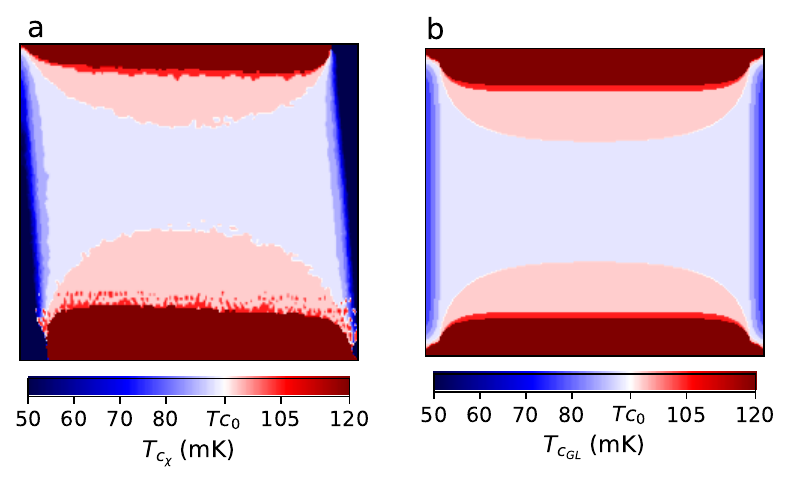}
    \caption{(a) $T_{c_{\chi}}$ map of the 100x100$\mu$m square device obtained from the image series in Fig.\ref{fig:100um_suscep_imgs}. We again observe a 'hourglass' feature of the local $T_{c_{\chi}}$ similar to what is described in the main text from the combined proximity and inverse proximity effects from the leads and banks respectively. Much of the device appears light blue which indicates a $T_{c_{\chi}} < T_{c_0}$. This is likely due to the narrow transition width in this device and the comparably coarse temperature spacing in the image series. The $\delta M(T)$ at the center of the device (Fig.3(a) in the main text) is measured with finer temperature spacing and suggests that $T_{c_{\chi}}$ at the center of the device is slightly higher than $T_{c_0}$. (b)  $T_{c_{GL}}$ map calculated by solving the GL equations as described in Supp. Sec.\ref{sec:GL}. Using the same GL parameters as in the main text, the map captures the key features of the measured $T_{c_{\chi}}$ map.}
    \label{100um_Tc_map}
\end{figure}

\newpage
\newpage
\stepcounter{somecounter}
\section{Supplementary Note \thesomecounter: Determining $T_{c_0}$}\label{sec:Tc0}

\begin{figure}[htbp]
    \includegraphics[width=1\linewidth]{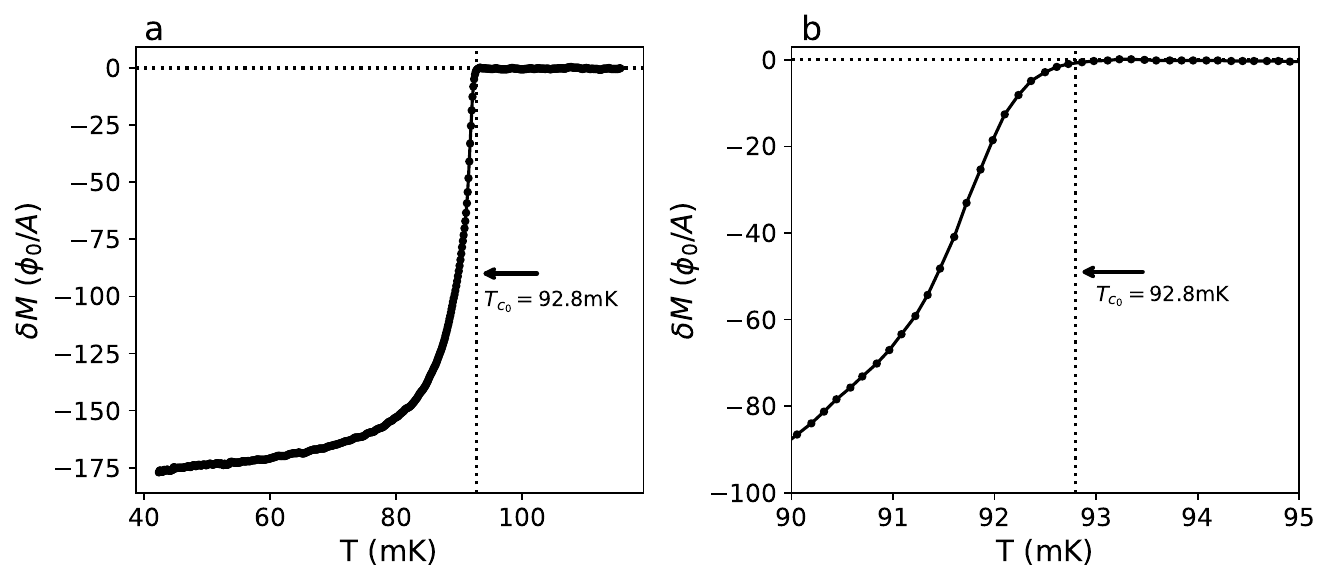}
    \caption{We determine $T_{c_0}$ of the bare MoAu bilayer by measuring the DR $\delta M(T)$ and finding the onset temperature at which $\delta M(T)<0$. (a) $\delta M(T)$ plotted over the full temperature range (same as Fig.3 in the main text). (b) Same data as (a), but zoomed in close to $T_{c_0}$. Vertical dashed lines in (a) and (b) correspond to $T_{c_0} =$ \SI{92.8}{\milli\kelvin}. We note that in addition to thermometry calibration, there is some some uncertainty in the exact determination of $T_{c_0}$ due to the noise floor of the measurement, which we estimate to be $\sim$\SI{0.2}{\milli\kelvin} corresponding to the approximate temperature spacing in the measurement. However, the temperature stability during imaging is a few 10s of \SI{}{\micro\kelvin}, so relative changes in temperature from $T_{c_0}$ during successive images is known more precisely with an uncertainty $<$\SI{0.1}{\milli\kelvin}.} 
    \label{fig:Tc0}
\end{figure}

\newpage
\newpage

\bibliography{bib_supp.bib}
\makeatletter\@input{xx2.tex}\makeatother